\documentclass[12pt,a4paper]{report}
\usepackage[utf8]{inputenc}
\usepackage{amsmath, amssymb}
\usepackage{geometry}
\usepackage{hyperref}
\usepackage{graphicx}
\usepackage{algpseudocode}
\usepackage{algorithm}
\usepackage{float}
\usepackage{booktabs}
\usepackage{listings}
\usepackage{xcolor}
\newtheorem{definition}{Definition}

\begin{document}

\begin{titlepage}
    \centering
    \vspace*{2cm}
    {\LARGE \textbf{Analysis and Comparison of Known and Randomly Generated S-boxes for Block Ciphers}\par}
    \vspace{1.5cm}
    {\Large Student: James Kim\par}
    \vspace{0.5cm}
    {\Large Supervisor: Dr. Graham Taylor Russell\par}
    \vspace{0.5cm}
    {\Large MSc Cryptography\par}
    \vspace{0.5cm}
    {\Large London Metropolitan University\par}
    \vfill
    {\Large August 2025\par}
\end{titlepage}

\begin{abstract}
Mathematically constructed S-boxes arise from algebraic structures and finite field theory to ensure strong, provable cryptographic properties. These mathematically grounded constructions allow for generation of thousands of S-Boxes with high nonlinearity, APN properties, and balanced avalanche characteristics, unlike fully random methods, which lack such theoretical guarantees in exchange for low complexity and more varied results. In this work, we compare mathematically constructed constructions with randomly generated ones to evaluate the relative weakness of the latter. We also establish an average measure of performance for randomly generated permutations, as well as random with forced cycle constraints, and compare them to well-established designs in a simple SPN setting.
\end{abstract}


\tableofcontents
\listoffigures
\listoftables
\listofalgorithms
\newpage

\chapter{Theoretical Background}
\section{Notations and Abbreviations Used}
$|A|$ - Cardinal of set A \newline
$a\oplus b$ - Bitwise XOR of a and b\newline
$\mathbf{GF(2^n)}$ - Galois (Finite) field with $2^n$ elements\newline
$6fa1_{16}$ - Hexadecimal representation \newline
$0101_{2}$ - Binary representation \newline
P-box - Permutation Box \newline
S-box - Substitution Box \newline
SPN - Substitution-Permutation Network \newline
DDT - Differential Distribution Table \newline
LAT - Linear Approximation Table \newline
MAX BIAS - Maximum value in the LAT \newline
PN - Perfect Nonlinear \newline
APN - Almost-Perfect Nonlinear \newline
$\Delta$ - Differential Uniformity \newline
EA - Extended Affine equivalence \newline
CCZ - Carlet Charpin Zinoviev equivalence \newline
SAC - Strict Avalanche Criterion \newline
BIC - Bit Independence Criterion \newline
AI - Algebraic Immunity \newline
PRNG - Pseudo-Random Number Generator

\newpage

\section{Introduction to Cryptography and Symmetric Block Ciphers}
Since the early days of recorded history there has been evidence of a need for secure communication. The broad tool that is cryptography today, even before it was formalised by Claude Shannon in his then-classified work "A Mathematical Theory of Cryptography" \cite{shannon} was used to provide such security. Early ciphers were built on substitution — replacing symbols with others according to some rule, such as in the Caesar or Playfair ciphers — or transposition, where only the order of symbols was altered, as in the Rail Fence cipher. These early systems were relatively simple but already captured the two fundamental building blocks of modern symmetric cryptography: substitution and permutation. \newline \newline
Modern block ciphers generalise these principles into the Substitution–Permutation Network (SPN) structure, which is widely used in practical cryptographic standards such as the Advanced Encryption Standard (AES). In an SPN, the plaintext input is divided into blocks, and each block passes through multiple rounds of transformation. Each round applies a nonlinear substitution layer (S-box), a linear permutation layer (P-box), and a key addition step. The design of these components is entirely public, with the key being the only secret, in accordance with Kerckhoffs’ principle. The substitution layer is the sole source of nonlinearity in the cipher, and therefore the properties of the S-boxes largely determine the cipher’s resistance to powerful attacks such as differential and linear cryptanalysis. For this reason, the study and construction of secure S-boxes remains one of the most active areas in symmetric cryptography.

\section{The Importance of S-boxes}
A block cipher is an algorithm that operates on states of fixed length instead the whole plaintext input or a single character from the plaintext. In practice, it is used within a mode of operation initialised with an Initialisation Vector (IV), which provides additional security, such as Cipher Block Chaining (CBC). The "symmetric" aspect indicates that the cipher uses the same key to encrypt and decrypt information. A Substitution-Permutation Network uses substitution and permutation layers in succession to achieve encryption, as opposed to a Feistel scheme, which often splits the block further and changes the state part-wise. The most important difference is that SPNs require bijective functions and have a simple decryption procedure which consists of applying the inverse functions in a reversed order, while the functions used in Feistel networks are not necessarily bijective, and the decryption is not as straightforward.
In modern symmetric ciphers, S-boxes are functions $S:\mathbf{GF(2^n)}\rightarrow\mathbf{GF(2^m)}$, where $\mathbf{GF(2^n)}$ is the Galois field with $2^n$ elements. This choice of these finite fields is well thought out as it builds on the idea of computers using bits — objects that can have two states — hence a field with $2^n$ elements is natural. The elements can be represented as polynomials of degree at most $n-1$ with coefficients in $\mathbb{Z}/2\mathbb{Z}$. In this field, the addition is simply the addition of polynomials, equivalent to bitwise XOR of the vectors formed from the coefficients. Multiplication is the polynomial multiplication modulo an irreducible polynomial of degree $n$. Detailed descriptions of finite fields used in cryptogrphy are found in \cite{ffields} chapters 2 and 9. A well established example is the field $\mathbf{GF(2^8)}$ used with the irreducible $x^8+x^4+x^3+x+1$ in the Advanced Encryption Standard \cite{aes}. The first S-boxes were designed and used by an IBM research group led by Horst Feistel in the Lucifer cipher \cite{lucifer}. Its successor, Data Encryption Standard, uses 8 different S-boxes in parallel that convert 6-bit inputs to 4-bit outputs. The DES S-boxes were extensively studied and shaped modern cryptanalysis. The most notable studies are of Biham and Shamir \cite{shamirdes}, who underlined the strong choice of S-boxes in DES with respect to differential cryptanalysis and of Matsui \cite{matsui}, who first proposed the ideas of a linear attack. Most of the modern ciphers use $8\times8$ or $4\times 4$ bit S-boxes. AES \cite{aes} and the comparable Camellia \cite{camellia} use 8x8 s-boxes while ciphers designed for lightweight applications, such as Serpent \cite{serpent}, PRESENT \cite{present} and Noekeon \cite{noekeon} use one or more 4x4-bit S-boxes in parallel for hardware efficiency.

\begin{figure}[h!]
    \centering
    \includegraphics[width=0.6\textwidth]{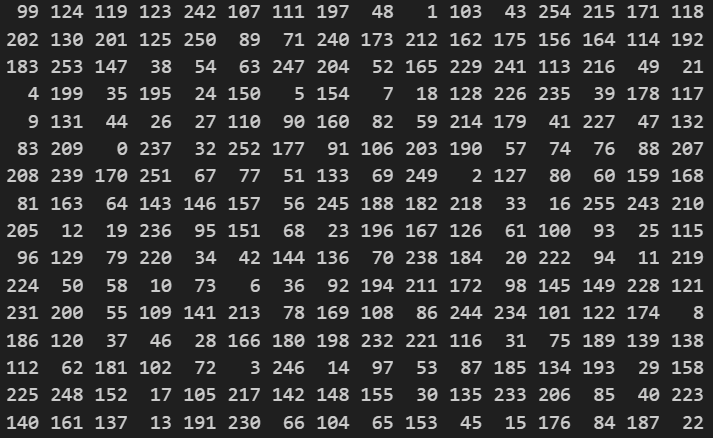}
    \caption{Rijndael S-box used in AES in decimal representation}
    \label{aessbox}
\end{figure}

\section{Motivation}
In this project we compare mathematically constructed $8\times8$ S-boxes with randomly generated ones in order to evaluate the relative weakness of the latter. Structured designs have become the standard and can provide better measure guarantees, while random ones may introduce unpredictable patterns with low complexity cost. We also establish an average measure of performance for randomly generated permutations, as well as random with forced cycle constraints. The results here are by all means reproducible and could serve as a baseline for improved random approaches such as chaos equations. In this way, the work serves both as a guide to the desired cryptographic properties and as a reference point for future random generation.

\section{Desirable Properties}
Claude Shannon (\cite{shannon} chapter 36) introduced confusion along diffusion as an essential principle in cipher design. In modern ciphers, confusion refers to making every bit in the plaintext depend on multiple bits in the key, making the relationship between them as complex as possible. This is now a property directly enforced by nonlinear layers - S-boxes. The natural problem that arises is how is the quality of such layers measured. Since the development of DES, several metrics have been developed, in order to underline resistance against particular mathematical attacks. The following are metrics commonly presented when analysing S-boxes. We spend more time describing differential uniformity and special cases here as it is commonly regarded as the first significant metric.

\subsection{Differential Uniformity}
\begin{figure}[h!] 
    \centering
    \includegraphics[width=0.6\textwidth]{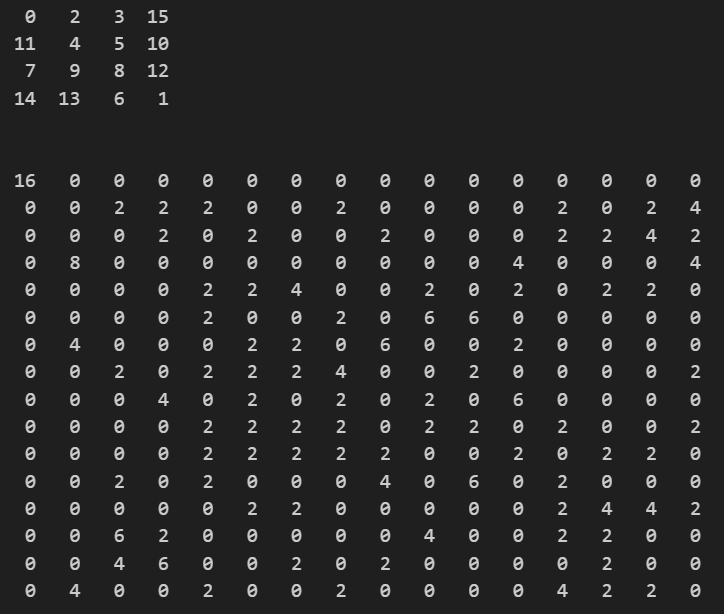}
    \caption{a 4x4 S-box and its DDT}
    \label{smallsboxddt}
\end{figure}

\begin{figure}[h!] 
    \centering
    \includegraphics[width=0.6\textwidth]{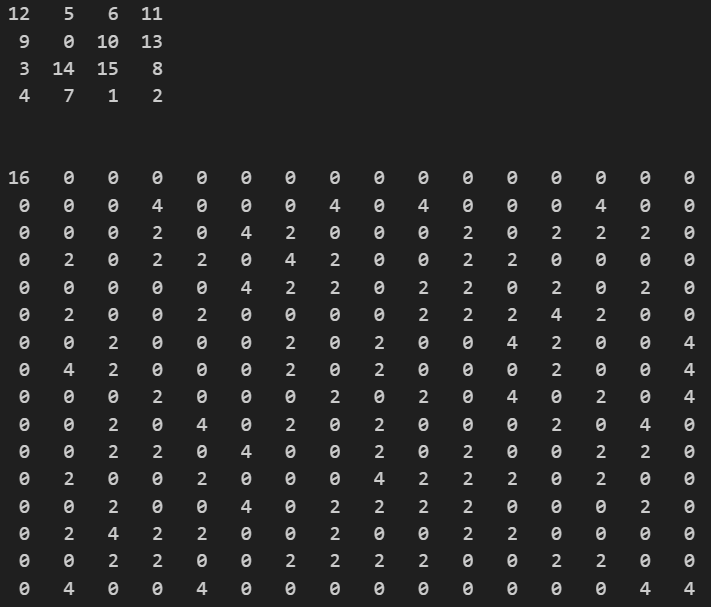}
    \caption{PRESENT S-box and its DDT}
    \label{smallsboxpresent}
\end{figure}

Differential uniformity is a measure of the S-box resistance to differential attacks that was introduced by Kaisa Nyberg \cite{nyberg}. We first define the difference distribution table.
\begin{definition}
    Let S be a function $S:\mathbf{GF(2^n)}\rightarrow\mathbf{GF(2^n)}$ and $a, b\in\mathbf{GF(2^n)}$. The difference distribution table $\mathrm{DDT}$ is a $2^n\times2^n$ matrix with values:
\begin{equation}
    \mathrm{DDT}[a,b]=|\{x\in\mathbf{GF(2^n)}|S(x)\oplus S(x\oplus a)=b\}|
\end{equation}
\end{definition}
The value at coordinate $[0, 0]$\footnote{Coordinates start at 0, so 0th row means the first row and so on. This numbering is natural as the decimal representations of the numbers in the finite fields go from 0 to $2^n-1$ inclusive.} in the table will always be $2^n$, i.e. the number of pairs $(x, x)$ such that $S(x)\oplus S(x) =0$, which is true for all $2^n$ values since the field characteristic is 2. The other elements of the row $a = 0$ will be $0$ as $S(x)\oplus S(x) =b$ is true iff $b = 0$. Hence only nonzero values of a are relevant.
\begin{definition}
The differential uniformity is then:
\begin{equation}
\Delta(S)=\mathrm{max}_{a\neq0,b}\mathrm{DDT[a,b]}
\end{equation}
\end{definition}
The values in the DDT will always be even, since if $x_0$ is a solution to $S(x) \oplus S(x\oplus a)=b$ then $x_0 \oplus a$ will also be a solution. If there are many solutions, i.e. the values in DDT are big, it translates to more pairs of different inputs. Hence when designing an S-box, the value of $\Delta$ and the number of entries in the DDT equal to $\Delta$ should be minimised.\newline
Figure \ref{smallsboxddt} shows a 4x4 S-box and its DDT. In this paper, all S-boxes are represented as $n\times n$ matrices of integers base 10 or base 16. For example, the function in \ref{smallsboxddt} is read as $f(0) = 0, f(1)=2, ... ,f(15)=1$. From this DDT, we can calculate the differential uniformity $\Delta=8$, that is, for $a = 3$ and $b=1$ the equation $f(x)\oplus f(x\oplus 3)=1$ has 8 solutions in $x$. So this function would not be particularly strong against a differential attack. Figure \ref{smallsboxpresent} shows the S-box used in the PRESENT \cite{present} cipher. It has differential uniformity 4.

\subsubsection{Almost Perfect Nonlinear (APN) Functions}

\begin{definition}
    An APN function f over $\mathbf{GF(2^n)}$ is a function f such that $\Delta(f)=2$
\end{definition}

Almost perfect nonlinear APN functions are optimal with respect to differential uniformity, having $\Delta$ = 2 and were studied by Nyberg in \cite{nyberg2}. They are of exceptional importance due to maximal differential resistance, however they unfortunately do not exhibit a simple general pattern. Several families of APN monomials have been discovered and presented so far from the works of Gold \cite{gold}, Kasami \cite{kasami} and Dobbertin \cite{dobbertinniho}, \cite{dobbertinwelch}.

\begin{table}[h!]
    \centering
    \begin{tabular}{ccc}
        Name & Function &  Condition\\
        \hline
        Gold & $x^{{2^i}+1}$ & gcd($i, n$)$=1$ \\
        Kasami & $x^{{2^{2i}}-2^i+1}$ & gcd($i, n$)$=1$ \\
        Welch & $x^{{2^t}+3}$ & $n=2t+1$\\
        Niho-1 & $x^{2^t+2^{\frac{t}{2}}-1}$ & $n=2t+1$, t even\\
        Niho-2 & $x^{2^t+2^{\frac{3t+1}{2}}-1}$ & $n=2t+1$, t odd\\
        Dobbertin & $x^{{2^{4i}}+2^{3i}+2^{2i}+2^{i}-1}$ & $n=5i$\\
        Inverse & $x^{{2^{2t}}-1}$ & $n=2t+1$\\
    \end{tabular}
    \caption{APN Monomials}
    \label{tab:placeholder}
\end{table}

The "Almost" in APN underlines that they are not perfect nonlinear, but the next best alternative. This is because there are no PN functions over fields of characteristic 2, as every solution $x_0$ always produces a different $x_0+a$ in the DDT, so a value $\Delta$ = 1 is impossible in the fields $\mathbf{GF(2^n)}$. While looking for functions with a desired DU (such as APN) it is important to describe what transformations preserve differential uniformity among other properties.

\subsubsection{Extended Affine and Carlet Charpin Zinoviev equivalence}
\begin{definition}
    A permutation of the elements $\mathbf{GF(2^n)}$ is a function $P:\mathbf{GF(2^n)}\rightarrow\mathbf{GF(2^n)}$ such that $P(x)$ is injective or surjective.
\end{definition}
Injectivity $\iff$ surjectivity in a finite field. \cite{ffields}
\begin{definition}
A linear map over $\mathbf{GF(2^n)}$ is a function $L:\mathbf{GF(2^n)}\rightarrow\mathbf{GF(2^n)}$ such that:
\begin{equation}
    L(x\oplus y)=L(x)\oplus L(y), \quad\forall x,y\in \mathbf{GF(2^n)}
\end{equation}
\end{definition}
This is equivalent to multiplication by an invertible $n\times n$ matrix with binary elements.

\begin{definition}
An affine map over $\mathbf{GF(2^n)}$ is a function of the form:
\begin{equation}
    A(x) = L(x)+a,\quad a\in\mathbf{GF(2^n)}
\end{equation}

\end{definition}
Where $L(x)$ is linear.
\begin{definition}
Two functions $S, T:\mathbf{GF(2^n)}\rightarrow\mathbf{GF(2^n)}$ are called EA-equivalent if there exist affine permutations $A_1$ and $A_2$ on $\mathbf{GF(2^n)}$ and affine (not necessarily a permutation) $A_3$ such that:
\begin{equation}
    T(x) =A_1\circ S\circ A_2(x)\oplus A_3(x)
\end{equation}
\end{definition}
This means that if T can be obtained by a relabelling of the inputs and outputs along XORing with a linear offset it has the same structure with respect to cryptographic properties so it is essentially the "same" function as S. A broader such relation introduced by Carlet, Charpin and Zinoviev in \cite{ccz} is the CCZ equivalence.
\begin{definition}
    Two functions $S, T:\mathbf{GF(2^n)}\rightarrow\mathbf{GF(2^n)}$ are called CCZ-equivalent if there exists an affine permutation $A:\mathbf{GF(2^n)}\times\mathbf{GF(2^n)}\rightarrow\mathbf{GF(2^n)}\times\mathbf{GF(2^n)}$ such that:
\begin{equation}
    A(\mathrm{Graph}(S)) = \mathrm{Graph}(g)
\end{equation}
$\mathrm{with}$
\begin{equation}
    \mathrm{Graph}(f)=\{(x,f(x))|x\in\mathbf{GF(2^n)}\}
\end{equation}
\end{definition}
The graph can be viewed simply as a set of ordered pairs. If the graph of a function can be transformed invertibly and respecting the affine condition, to the graph of another, they are essentially the same function with respect to input-output relations. As proved by Carlet, Charpin and Zinoviev, CCZ equivalent functions have the same differential uniformity and nonlinearity. Moreover they proved that CCZ equivalence is strictly more general than EA equivalence \cite{ccz}, with results also clearly restated in Carlet's chapter on Boolean functions in \cite{ffields}.\newline
In the newer works that aim to classify APN functions, all classifications are done up to CCZ equivalence. In addition to the monomial forms in Table 1.1, there are many other known APN families. The work of Kangquan Li and Nikolay Kaleyski \cite{kangquan} presents two infinite families and lists the known ones of polynomial form up to CCZ equivalence.

\subsubsection{APN permutations and small even dimensions 4, 6, 8}
The elements of $\mathbf{GF(2^n)}$ can be represented as n-bit strings describing coefficients of the corresponding polynomial and computers are designed based on the duality of a bit's state - 0 or 1. This naturally points us to even choices of dimension when building strong functions and even more to powers of 2. SPN networks aditionally require all their transformations to be invertible for decryption. This is why the focus of many works has been on finding APN functions that are bijections in even dimensions. It has been conjectured that there are no APN permutations in even dimensions. Computational and theoretical proofs \cite{hou} for $n=4$ supported this conjecture. However, Browning, Dillon, McQuistan and Wolfe \cite{dillon1} provided a counterxample in dimension 6, shown in \ref{dillperm}, after applying many restrictions including on objects defined as dual codes from the graph of $f$ and setting $f(0)=0$. In their paper, the main turning point is disproving a conjecture by Hou, stated in \cite{hou}, for dimension 6. The newly discovered APN permutation was then naturally deconstructed, and attempts were made to generalise it for other even dimensions \cite{dillongen}. For n = 8 and above even dimensions it remains an open problem whether there are any APN permutations. Hence the best differential uniformity achievable in this case is 4, as used in AES.  Advancements on this problem include looking at classes of APN functions and checking their bijectivity. Beierle, Brinkmann, and Leander designed a recursive tree-search algorithm specifically for finding quadratic APN functions with linear self-equivalences in small dimensions. Their search unearthed 12,921 new quadratic APN functions in dimension 8 up to CCZ equivalence, though none were confirmed to be permutations.

\begin{table}[h]
    \centering
    \begin{tabular}{cccccccc}
        0 & 54 & 48 & 13 & 15 & 18 & 53 & 35\\
        25 & 63 & 45 & 52 & 3 & 20 & 41 & 33\\
        59 & 36 & 2 & 34 & 10 & 8 & 57 & 37\\
        60 & 19 & 42 & 14 & 50 & 26 & 58 & 24\\
        39 & 27 & 21 & 17 & 16 & 29 & 1 & 62\\
        47 & 40 & 51 & 56 & 7 & 43 & 44 & 38\\
        31 & 11 & 4 & 28 & 61 & 46 & 5 & 49\\
        9 & 6 & 23 & 32 & 30 & 12 & 55 & 22\\
    \end{tabular}
    \caption{APN permutation in dimension $n=6$}
    \label{dillperm}
\end{table}

\subsection{Nonlinearity and Bias}
Nonlinearity measures how far is a given function, in our case a vectorial Boolean function $S:\mathbf{GF(2^n)}\rightarrow\mathbf{GF(2^n)}$ is from every affine function. High nonlinearity is crucial as it represents the resistance against linear cryptanalysis introduced by Matsui \cite{matsui}. While differential cryptanalysis exploits uneven propagation of differences through the cipher, linear cryptanalysis exploits possible approximation of the S-box by an affine function. We now give mathematical definitions of nonlinearity and prerequisite concepts.
\begin{definition}
    The Walsh transform of a function $f:\mathbf{GF(2^n)}\rightarrow\mathbf{GF(2)}$ is the value
\begin{equation}
    W_f(a)=\sum_{x\in \mathbf{GF(2^n)}}{(-1)^{f(x)\oplus a\cdot x}}
\end{equation}
where the sum is the usual addition (not XOR), and the set
\begin{equation}
    \{W_f(a)|a\in\mathbf{GF(2^n)}\} 
\end{equation}
is the Walsh spectrum of $f$.
\end{definition}
The dot product here is the binary dot product i.e. $x\cdot y=x_0\& y_0\oplus x_1\&y_1\oplus \cdots \oplus x_{n-1}\&y_{n-1}$ where $x_k,y_k$ are the bits of $x$ and $y$. The Walsh transform is a discrete Fourier transform used in $\mathbf{GF(2^n)}$ that measures correlation or similarity to all linear functions. Therefore we want the modulus $|W_f(a)|$ to be small. An ideal value of 0 means, across all inputs, that the function produces the same output as a linear approximation with perfect probability $\frac{1}{2}$. For an $n\times n$ S-box, a vectorial Boolean function, we need to consider all contained functions $f:\mathbf{GF(2^n)}\rightarrow\mathbf{GF(2)}$ and we populate a table with all Walsh transform values.
\begin{definition}
    Let S be a function $S:\mathbf{GF(2^n)}\rightarrow\mathbf{GF(2^n)}$ and $a, b\in\mathbf{GF(2^n)}$. The linear approximation table $\mathrm{LAT}$ is a $2^n\times 2^n$ matrix where
    \begin{equation}
        \mathrm{LAT}[a,b] =\sum_{x\in \mathbf{GF(2^n)}}{(-1)^{b\cdot S(x)\oplus a\cdot x}}
    \end{equation}
\end{definition}
As with DDT, the coordinate LAT[0,0] will always be equal to $2^n$ so we ignore it in further calculations. The maximum bias (or maximum correlation) is the maximum absolute value in the LAT, excluding the first row and column, and it emphasizes the weakest bits, i.e. where a linear attack would start from. This value is sometimes normalised by dividing by $2^n$.
\begin{definition}
    Let f be a function $f:\mathbf{GF(2^n)}\rightarrow\mathbf{GF(2)}$. The nonlinearity of $f$ is
    \begin{equation}
        NL_f=2^{n-1}-\frac{1}{2}\max_{a\in \mathbf{GF(2^n)}}|W_f(a)|
    \end{equation}
\end{definition}
The nonlinearity of an S-box will then be given by the lowest nonlinearity across all its contained Boolean functions. We often also report element-wise nonlinearity, max and average. The theoretical upper bound of nonlinearity is $2^{n-1}-\frac{1}{2}2^{n/2}$. The functions that achieve this condition form the class of Bent functions. Bent functions were extensively studied by Rothaus \cite{rothaus} and in \cite{canteaut}, \cite{carletnon} and are, alongside APN functions, of great interest to cryptography for their properties. However, Bent functions cannot be used directly as components of S-boxes, because they are not balanced i.e. their values are skewed more to 0 or 1, and hence they cannot form permutations. The process of finding good nonlinearity values involves starting with a bent-based S-box and changing outputs slightly, while checking the value NL at each step. For example the AES S-box has nonlinearity 112, close to the maximum of 120.

\subsection{Strict Avalanche Criterion}
The avalanche effect was observed by Feistel \cite{feistel1} and completely described by Webster and Tavares \cite{webster}. The strict avalanche criterion states that a single bit-flip in the input should produce a flip in the each of the output bits with probability $\frac{1}{2}$. This is the ideal, perfect case, which is not achieved in practice. Instead we consider the Distance from SAC. This measures how far is a function from achieving SAC, either by a probabilistic deviation from $\frac{1}{2}$ or the actual number of bit-flip difference from $2^{n-1}$.
\begin{definition}
    Let $f:\mathbf{GF(2^n)}\rightarrow\mathbf{GF(2)}$ and the set 
    \begin{equation}
        A_i=\{x|f(x)\oplus f(x\oplus 2^i)=1\}.
    \end{equation}
    Where i takes values from 0 to $n-1$, i.e. XORing with $2^i$ represents all single bit flips.
    The DSAC of $f$ is the value 
    \begin{equation}
        DSAC=\max_i||A_i|-2^{n-1}|
    \end{equation}
\end{definition}
The result is sometimes divided by $2^n$ for normalisation.
For an S-box we often report the maximum and the mean DSAC across all component functions. The value of the DSAC should be minimised when designing S-boxes, this translates to less deviation from half bit-flips. For example, the AES S-box has max DSAC = 0.0625\footnote{All values cited here were obtained using the Python code in sboxes\_6.ipynb}, or 16 in non-normalised form, which reveals a worst-case deviation of 16 bit flips from 128 out of 256, and mean DSAC = 0.0263671875.

\subsection{Bit Independence Criterion}
The Bit Independence Criterion is another avalanche-based measure similar to SAC, introduced in \cite{webster}. Instead of considering single flips, we measure if a flip in the input creates a dependence in the flips of two bits form the output. In other words, how "dependent" are two output bits from one another.
\begin{definition}
    Let $f:\mathbf{GF(2^n)}\rightarrow\mathbf{GF(2)}$, input bit $i$, output bits $j, k$. Define the set
    \begin{equation}
        A_{i,j,k}=\{x|f_i(x)\oplus f_i(x\oplus 2^i)=1\quad and\quad x|f_j(x)\oplus f_j(x\oplus 2^i)=1\}.
    \end{equation}
    and the probability
    \begin{equation}
        p_{i,j,k}=\frac{|A_{i,j,k}|}{2^n}
    \end{equation}
\end{definition}
BIC is achieved when the bits are completely independent, so the simultaneous flip should happen with probability $\frac{1}{2}\cdot \frac{1}{2}=\frac{1}{4}$
\begin{definition}
The deviation from BIC for input $i$ and output pair $j,k$ is
\begin{equation}
    D_{i,j,k}=|\frac{1}{4}-p_{i,j,k}|
\end{equation}
and the maximum Distance from BIC per S-box is:
\begin{equation}
    DBIC=\max_{i\in\{0, ...,n-1\}}(\max_{j\neq k\quad j,k\in\{0, ...,n-1\}}D_{i,j,k})
\end{equation}
\end{definition}
As it is the case with DSAC, a low BIC indicates a stronger S-box, meaning fewer pairs of output bits will be correlated. The max DBIC in the AES S-box is 0.0703125.

\subsection{Cycle Structure}
As our S-boxes need to be permutations of $\mathbf{GF(2^n)}$, we can examine the number of their cycles and their size as well as fixed and opposite fixed points. Fixed points are values $x$ such that $S(x)=x$, opposite fixed points have $S(x) = \neg x$ where $\neg x\oplus x=2^n-1$ or $11...1_2$. The latter create potential weaknesses, as the difference $S(x)\oplus x=11...1_2$ is very predictable and could be exploited in a differential attack. In SPNs however, the permutation layer shuffles and disrupts these entries, so a small number of them do not compromise security. The opposite fixed points correspond to 2-cycles $(x,\neg x)$. Having many short cycles in our permutation is not desirable because they can reduce mixing after multiple rounds.

\subsection{Algebraic Degree and Algebraic Immunity}
We first define the useful concept of Algebraic Normal Form (ANF).
\begin{definition}
    Let $f:\mathbf{GF(2^n)}\rightarrow\mathbf{GF(2)}$. The Algebraic Normal Form ANF $f$ is the unique polynomial representation:
    \begin{equation}
        f(x)=\bigoplus_{a\in\mathbf{GF(2^n)}}c_a\prod_{i=0}^{n-1}x_i^{a_i}
    \end{equation}
\end{definition}
The product used here represents the AND operator, or multiplication modulo 2, $x_i$ are the bits of $x$ and coefficients $c\in\{0,1\}$. In other words, the ANF is the Boolean function written in terms of ANDs and XORs of the bits from the input. This is derived in practice by constructing the truth table of $f$ over all possible inputs. Important measures that derive from this are the Algebraic Degree and Algebraic Immunity.
\begin{definition}
    The Algebraic Degree of $f$ is the highest degree across the monomial components XORed in the ANF of $f$.
\end{definition}
This can be also understood as the last nonzero coefficient $c$ from the ANF when using the above notation.

\begin{definition}
    The Algebraic Immunity of $f$ is the minimum possible degree of a nonzero function $g$ over the same field such that:
    \begin{equation}
        f(x)\cdot g(x)=0 \quad or \quad (f(x)\oplus1)\cdot g(x)=0, \quad \forall x\in\mathbf{GF(2^n)}.
    \end{equation}
\end{definition}
The algebraic immunity is the minimum degree of a function that cancels out $f$ or its complement. An $n\times n$ S-box is constructed from n Boolean functions operating on the bits $x_i$ of the input $S(x_0,x_1,...,x_{n-1})=(f_0(x_0,...,x_{n-1}),..,(f_{n-1}(x_0,...,x_{n-1})$. The Algebraic Degree and Algebraic Immunity of an S-box are then the maximum Degree and the minimum Immunity respectively across all component boolean functions. This measure is important as it assesses resistance against an algebraic attack. This cryptanalysis was first presented by Courtois and Pieprzyk in \cite{courtois}. In the method they propose, the cipher is represented as a system of multivariate polynomial equations over $\mathbf{GF(2)}$ and solving these equations viewing the key bits as unknowns. Albrecht and Cid \cite{cid} proposed a newer attack which mixes algebraic methods with earlier statistical approaches.

\subsection{Hamming Distance and Trace}

These are element-wise, not S-box wide operations but they will be useful later and are mentioned for completeness.
\begin{definition}
    The Hamming distance between elements $x, y\in\mathbf{GF(2^n)}$ is the number of 1 bits in the binary representation of $z =x\oplus y$
\end{definition}
In other words, how "far" in terms of needed single bit flips is $x$ from $y$. 
\begin{definition}
    The field trace of an element $x$ is the sum of all Galois conjugates of $x$, in our field ${GF(2^n)}$ this means:
    \begin{equation}
        Tr(x)=x^{2^0}\oplus x^{2^1} \oplus\cdots\oplus x^{2^{n-1}}
    \end{equation}
    The prime field of $\mathbf{GF(2^n)}$ is $\mathbf{GF(2)}$ so the trace will either be 0 or 1 in our case.
\end{definition}

\section{Methods of S-box Generation}
There are two main ways new strong S-boxes are discovered: through random methods or mathematical construction. The random approach is simple to implement and produces high-entropy, unpredictable functions. However, the good cryptographic properties mentioned above need to be checked for each generated S-box as there is no guarantee on good properties. Mathematically constructed S-boxes, on the other hand, are built using algebraic or number-theoretic structures over finite fields, such as the inverse function with an affine transformation used in the AES S-box or Bent functions which are then tweaked up to balanced outputs. These constructions allow designers to guarantee some of the properties across classes before testing the others.

\subsection{Pseudo-Random Number Generators}
In this approach, an S-box is generated by producing a random permutation of the set $\{0, ...,n-1\}$ using a cryptographically secure PRNG. The generated numbers are used to fill the S-box entries while ensuring bijectivity (each output value appears exactly once). Common methods include shuffling algorithms like the simple P algorithm proposed by Knuth in \cite{knuth} , seeded with a random key or entropy source. PRNG-based S-boxes are simple to implement and can provide high entropy, making them unpredictable. The main limitation is that, without additional constraints, the resulting S-box may have poor cryptographic properties such as high differential uniformity or low nonlinearity, so metric evaluation is necessary after generation. Therefore many tests may be needed in this case.

\subsection{Chaos Equations}
Chaotic systems, such as the logistic map or Henon map, are deterministic equations that exhibit pseudo-random behavior under certain parameters or after many iterations. To generate an S-box, one can iterate a chaotic equation to produce a sequence of numbers, then map these numbers to S-box entries while enforcing bijectivity. This approach is often limited to smaller sizes $(4 \times 4$ or $5\times 5)$. Chaos-based S-boxes exploit the sensitivity to initial conditions and complex dynamic behavior to create highly unpredictable S-boxes. They are a particularly interesting and quite recent tool used in this context. Examples include \cite{gaabouri} where authors used an enhanced form of the logistic map, or \cite{solami}, where authors used a 5-dimensional system of differential equations, shown in chapter 2 of their paper. However, care must be taken in turning resulted real numbers to integers and scaling the chaotic outputs, as poor parameter choices can reduce randomness or introduce patterns that weaken security.

\subsection{Random polynomials}
Polynomials over finite fields can also be used for pseudo-random S-box generation. By choosing exponents and that produce bijective mappings and iterating over all field elements, one can generate S-boxes with moderate control over the algebraic complexity. Certain exponents lead to almost perfect nonlinear (APN) functions, such as in the classes of functions presented in 1.4.1, reducing differential uniformity, while others can produce high algebraic degree. This method combines deterministic algebraic structure with pseudo-random-like output distribution, giving a balance between unpredictability and provable cryptographic properties.

\subsection{Mathematical Construction}
The mathematical construction of S-boxes leverages algebraic structures and finite field theory to ensure strong, provable cryptographic properties. The AES S-box was built using the inverse map $x\mapsto x^{-1}$ with respect to irreducible polynomial $x^8+x^4+x^3+x+1$. This has been further extended by Cui et al. in \cite{cui2}, by modifying the affine transformation added in generating AES S-box, resulting in a new S-box with better DSAC and algebraic complexity of the function. An even better result is constructed in \cite{nitaj1}, which further improves BIC and DSAC. Other creative methods use elliptic curves — one proposal employs elliptic curves of the form $y^2=x^3+n$ with integer $n$ defined over prime fields to generate S-boxes dynamically, providing efficient computation \cite{mordel}. Furthermore, algebraic structures known as semi-fields \cite{uga} have been used to form 'pseudo-extensions' that mimic the behavior of the finite field, allowing the generation of thousands of S boxes with high non-linearity, APN properties and balanced avalanche characteristics. These mathematically grounded constructions allow for S-box design with built-in resistance to linear and differential cryptanalysis, unlike fully random methods which lack such theoretical guarantees.

\chapter{Experiment}
The experiment consists of two stages of selection and evaluation of S-boxes. In the first stage, 100,000 random S-boxes are generated, and for each of the established cryptographic metrics: differential uniformity $\Delta$, maximum linear bias, SAC, BIC and nonlinearity. The best candidate is identified and the process is repeated under the constraint of fixed cycle structures, in which both the number and the length of cycles are fixed, in order to examine the influence of cycle decomposition on cryptographic properties. Statistical results are reported for all candidates. The DDTs and LATs of S-boxes obtained from both procedures are then presented using heatmaps to give a more visual insight of their properties. In the second part, the most favorable S-boxes are incorporated into a simple block cipher and subjected to a full-cipher avalanche evaluation. Comparative analysis is then conducted against the AES S-box and the S-boxes given in \cite{cui2} and \cite{solami}. The complete code implementations are provided in the supplementary files as "sboxes\_6.ipynb" (jupyter notebook), "julia\_search" (julia) for the S-box properties and "spn\_cipher.ipynb" for the full-cipher comparison. 

\section{Generating the S-boxes}
In this section, we describe the algorithms used to generate the best S-boxes with respect to each metric. Algorithm 1 produced the unconstrained S-boxes and Algorithm 2 produced the results with fixed cycle layout. $Base\_Value$ is set as the worst-case bound value for each metric (in practice, we set it as the value measured in the identity permutation). $Max\_Tries$ represents the search size, set at 100000 in the experiments. The $measure()$ function was used with metrics: Uniformity, Maximum Bias (max LAT value), Distance from SAC, Distance from BIC, Nonlinearity. In the fixed cycles case, we looked at five cycle structures: 64 4-cycles, 16 16-cycles, 4 64-cycles, one 256-cycle, and finally the uneven cycle structure of Rijndael S-box - with cycle sizes 59, 81, 87, 27, 2. All the 30 resulted S-boxes are provided in an additional text file, using decimal notation. The algebraic degree and immunity were not included here as their computational expense was overwhelming compared to the other metrics.
\begin{algorithm}

\caption{Random unconstrained}\label{randgen}
\begin{algorithmic}[1]
\State $Best\_metric \gets Base\_value$
\State $Best\_Sbox \gets Identity\_Permutation$
\State $Tries \gets 0$
\State $Mean\_Metric\_Value$
\While{$Tries <Max\_Tries$}
    \State $Current\_Sbox \gets Random\_Permutation(255)$
    \State $Tries \gets Tries +1$
    \State $Current\_Metric\_Value \gets measure(Current\_Sbox, Metric)$
    \If{$Current\_Metric\_Value$ is better than $Best\_Metric\_Value$}
        \State $Best\_Metric\_Value \gets Current\_Metric\_Value$
        \State $Best\_Sbox \gets Current\_Sbox$
    \EndIf
\EndWhile
\end{algorithmic}
\end{algorithm}

\begin{algorithm}
\caption{Random with given cycles}\label{randgenconst}
\begin{algorithmic}[1]
\State $Best\_metric \gets Base\_value$
\State $Best\_Sbox \gets Identity\_Permutation$
\State $Tries \gets 0$
\State $Mean\_Metric\_Value$
\While{$Tries <Max\_Tries$}
    $Cycles\_Filled \gets0$
    \While{$Cycles\_ Filled<Cycles\_Number$}
        \State $Cycle_{Cycles\_Filled}\gets Random\_Permutation_{255}(Cycle\_Size)$
        \Comment{Choose $Cycle\_Size$ numbers in range 0...255 that were not already used in previous cycles}
    \EndWhile
    \State Build $Current\_Sbox$ from $\{Cycle_0,...,Cycle_{Cycles\_Number}\}$
    \State $Tries \gets Tries +1$
    \State $Current\_Metric\_Value \gets measure(Current\_Sbox, Metric)$
    \State $Mean\_Metric \gets Mean\_Metric +(Current\_Metric\_Value)$
    \If{$Current\_Metric$ is better than $Best\_Metric$}
        \State $Best\_Metric\_Value \gets Current\_Metric\_Value$
        \State $Best\_Sbox \gets Current\_Sbox$
    \EndIf
\EndWhile
\end{algorithmic}
\end{algorithm}

\subsection{Running Specifications and Tools}

Single-S-box full measurements and full cipher calculations were made using the Python programming language, with the calculations of LAT and Nonlinearity being the most computationally heavy. For the random generation with single metric evaluation, we used the Julia language for much faster calculations over the 100000 S-boxes. The running time for each set of metrics was around two hours, again due to the many calculations involved in MAX BIAS and NL. The random number generator used therefore is xoshiro256++, as specified on the Julia documentation page.

\subsection{Measured properties and comparison}
The table \ref{tablemetrics} lists all metrics for each of the resulted S-boxes, compared to the S-boxes from \cite{cui2} and \cite{solami}. For each entry, the name is Optimisedmetric\_MethodOfGeneration, for example NL\_Rij\_Cyc has the best nonlinearity from the random generation with set Rijndael S-box cycles and DU\_4\_64 has the best Differential Uniformity from the S-boxes generated with 4 cycles of length 64. \newline The mean values represent the arithmetic mean of the generated set.

\begin{figure}[H]
    \centering
    \includegraphics[width=1.15\textwidth]{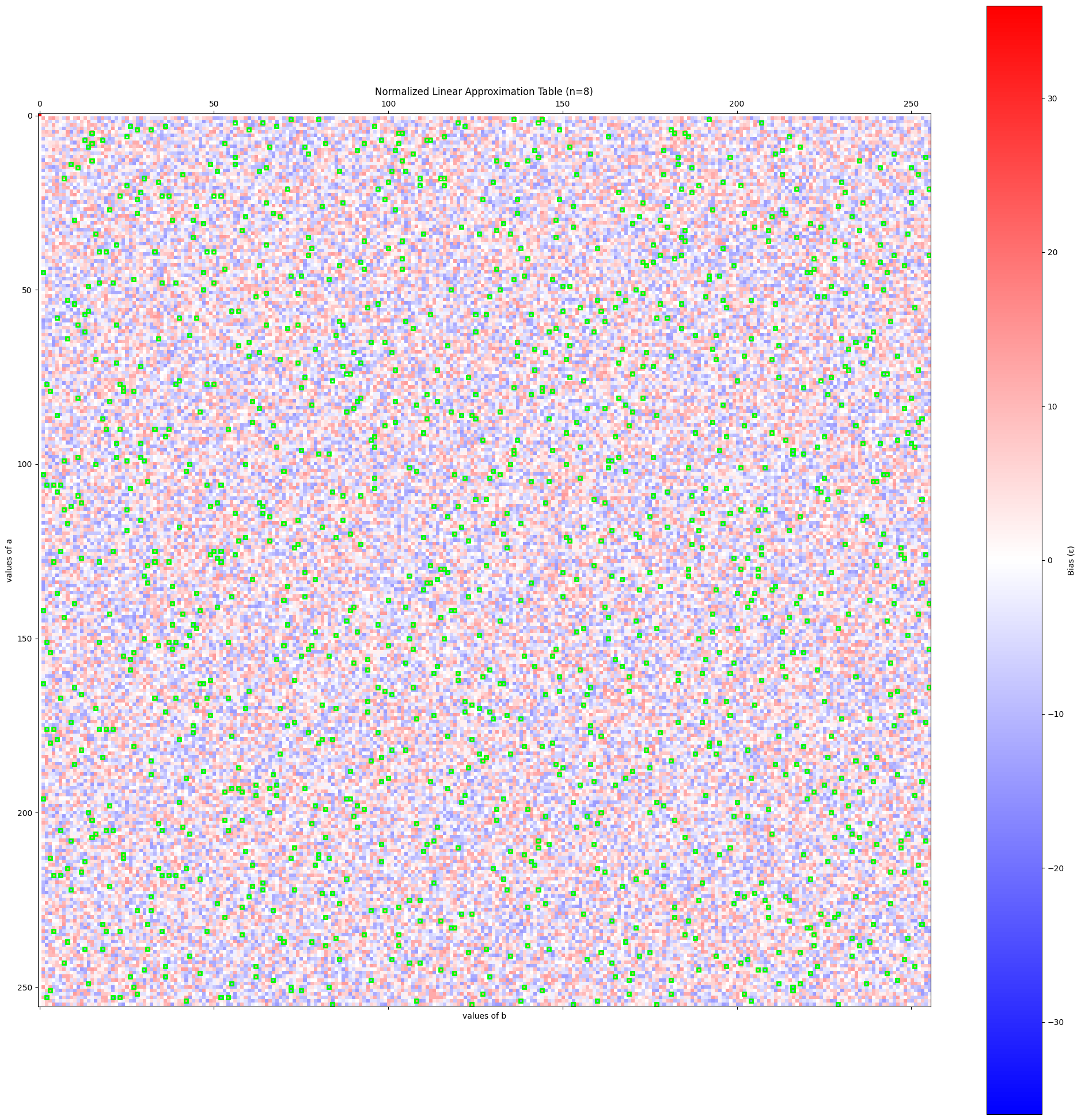}
    \caption{LAT visualisation of AES S-box}
    \label{aeslat}
\end{figure}

\begin{figure}[H]
    \centering
    \includegraphics[width=1.15\textwidth]{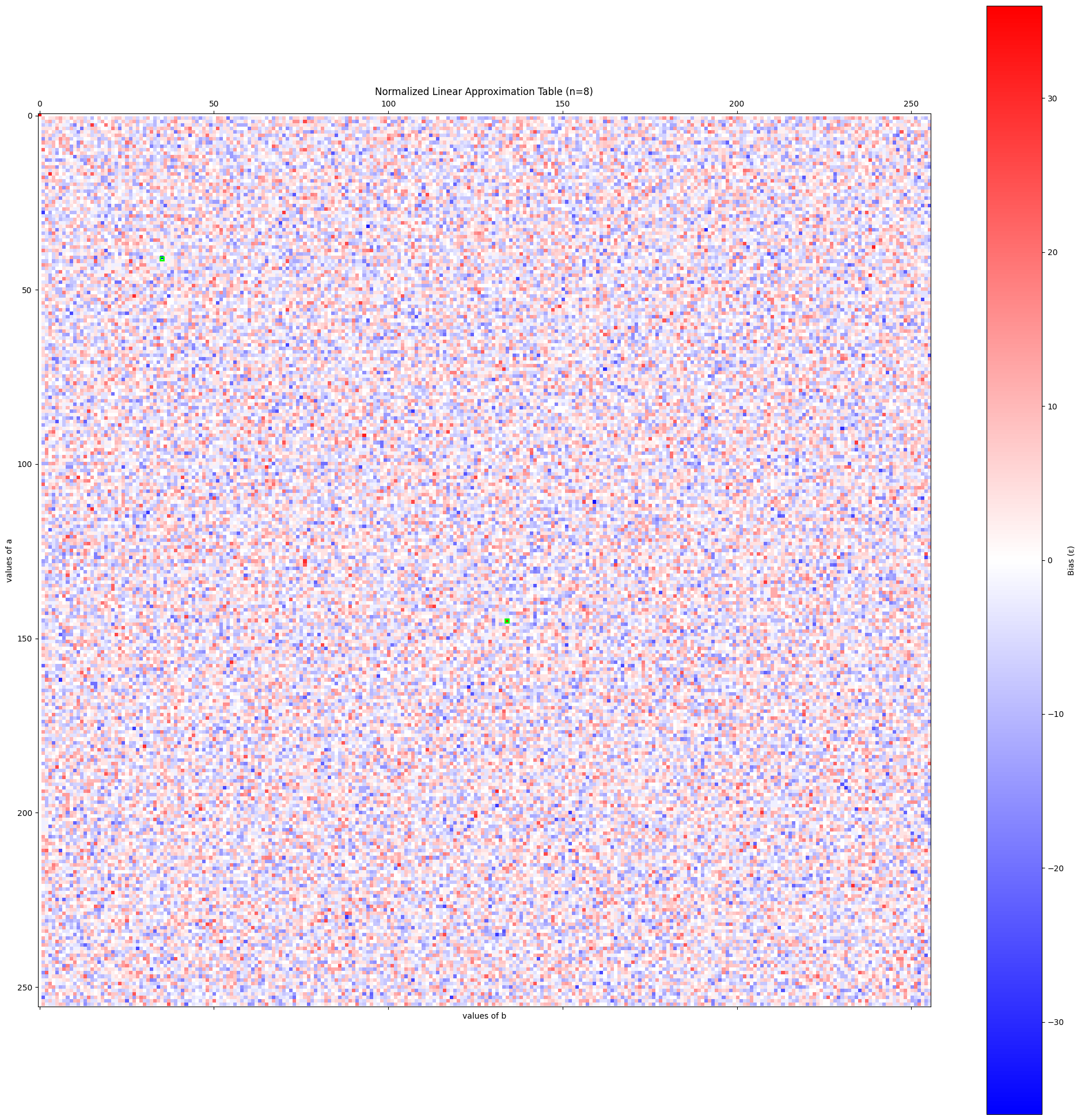}
    \caption{LAT visualisation of DBIC\_Rij\_Cyc}
    \label{rijcyc}
\end{figure}

\begin{figure}[H]
    \centering
    \includegraphics[width=1.15\textwidth]{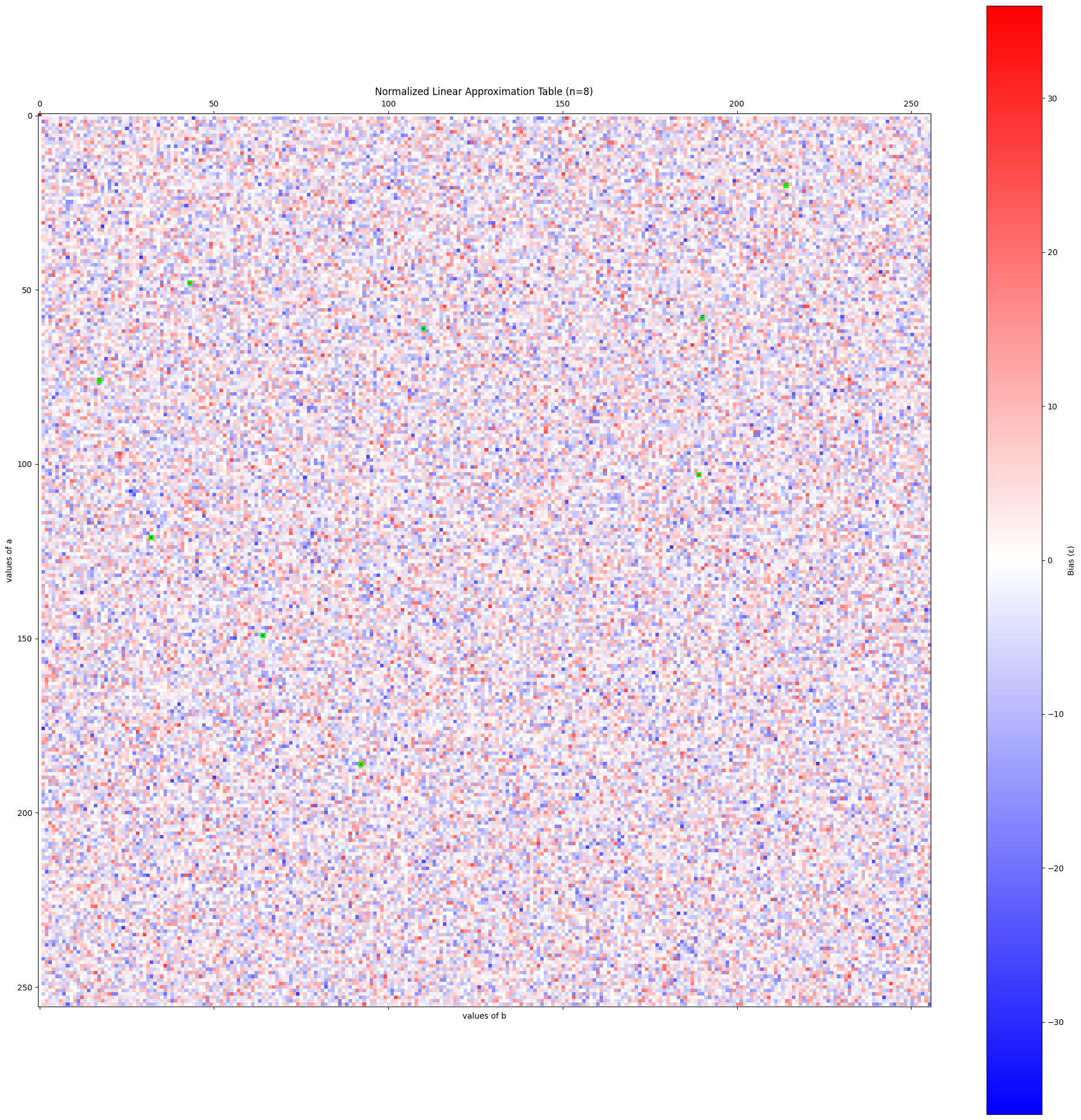}
    \caption{LAT visualisation of NL\_64\_4}
    \label{nl644}
\end{figure}

From the table we observe generally slightly better results in the balanced 16 16-cycles case, with better means in almost all properties. We also note that the DSAC of AES/CUI/SOLAMI was easily achieved in all searches, while the MAX BIAS and NL were the farthest. However, the mean values for each random search is very far from the constructed cases. This further emphasises the importance and superiority of mathematical constructions with respect to the measured properties. From this table, we consider two S-boxes that seem to stand out: \newline
DBIC\_Rij\_Cyc, with the best DBIC (better than AES)\newline
NL\_64\_4, best NL and MAX BIAS, worst  DSAC and DBIC \newline
Heatmaps of their LAT are shown \ref{rijcyc} \ref{nl644}, compared to AES \ref{aeslat}, to emphasise the differences in resistance against linear cryptanalysis. The scale shows bias values on the same scale -36 (blue) to 36 (red). This value was chosen as it is the worst value across the three S-boxes (MAX BIAS of DBIC\_Rij\_Cyc). Darker hues therefore indicate the weakest bits while whiter areas are closer to 0 (or perfect 0.5 probability). The green points indicate bits where the maximum value (worst) is achieved. We observe much lighter colours in the AES S-box as expected, as well as a more uniform distribution of the maximum bias entries, while the weaker S-boxes have much fewer and more isolated weak points. The resistance against a linear attack is given by the worst value, as the S-box and cipher structure are entirely public and a potential attacker could perform such analyses. There is also a slightly less obvious difference between DBIC\_Rij\_Cyc and NL\_64\_4, accounting for the difference 6 in their MAX BIAS. A list of all generated S-boxes from \ref{tablemetrics} with corresponding names is presented in Appendix \ref{appendix:sboxes}.

\begin{table}[H]
    \centering
    \begin{tabular}{|l|l|l|l|l|l|}
    \hline
        S-box & DU & MAX BIAS & DSAC & DBIC & NL \\ \hline
        AES & 4 & 16 & 0.0625 & 0.0703125 & 112 \\ \hline
        CUI \cite{cui2}& 4 & 16 & 0.0625 & 0.0625 & 112 \\ \hline
        SOLAMI \cite{solami}& 4 & 16 & 0.0625 & 0.0625 & 112 \\ \hline
        DU\_Random & 8 & 34 & 0.09375 & 0.1015625 & 94 \\ \hline
        MAX\_BIAS\_Random & 10 & 32 & 0.09375 & 0.1015625 & 96 \\ \hline
        DSAC\_Random & 10 & 34 & 0.0625 & 0.0859375 & 94 \\ \hline
        DBIC\_Random & 10 & 36 & 0.078125 & 0.0703125 & 92 \\ \hline
        NL\_Random & 12 & 30 & 0.125 & 0.125 & 98 \\ \hline
        MEAN\_Random & 11.34628 & 35.27799999 & 0.11395093 & 0.11382468 & 92.76599999 \\ \hline
        DU\_64\_4 & 8 & 36 & 0.15625 & 0.1328125 & 92 \\ \hline
        MAX BIAS\_64\_4 & 10 & 30 & 0.109375 & 0.1015625 & 98 \\ \hline
        DSAC\_64\_4 & 12 & 32 & 0.0625 & 0.1171875 & 96 \\ \hline
        DBIC\_64\_4 & 12 & 36 & 0.078125 & 0.0703125 & 92 \\ \hline
        NL\_64\_4 & 10 & 30 & 0.21875 & 0.171875 & 98 \\ \hline
        MEAN\_64\_4 & 11.34618 & 35.31399999 & 0.11397218 & 0.11385609 & 92.65799999 \\ \hline
        DU\_16\_16 & 8 & 36 & 0.15625 & 0.1328125 & 92 \\ \hline
        MAX BIAS\_16\_16 & 10 & 30 & 0.109375 & 0.1015625 & 98 \\ \hline
        DSAC\_16\_16 & 12 & 32 & 0.0625 & 0.1171875 & 96 \\ \hline
        DBIC\_16\_16 & 12 & 36 & 0.1171875 & 0.0703125 & 96 \\ \hline
        NL\_16\_16 & 12 & 36 & 0.140625 & 0.109375 & 98 \\ \hline
        MEAN\_16\_16 & 11.34322 & 35.31999999 & 0.11396734 & 0.11383835 & 92.77799999 \\ \hline
        DU\_4\_64 & 8 & 36 & 0.09375 & 0.09375 & 92 \\ \hline
        MAX BIAS\_4\_64 & 10 & 30 & 0.109375 & 0.1328125 & 98 \\ \hline
        DSAC\_4\_64 & 10 & 38 & 0.0625 & 0.0859375 & 90 \\ \hline
        DBIC\_4\_64 & 10 & 32 & 0.09375 & 0.0703125 & 96 \\ \hline
        NL\_4\_64 & 12 & 30 & 0.140625 & 0.109375 & 98 \\ \hline
        MEAN\_4\_64 & 11.33462 & 35.29599999 & 0.11399437 & 0.11387914 & 92.71799999 \\ \hline
        DU\_256\_1 & 10 & 34 & 0.125 & 0.140625 & 94 \\ \hline
        MAX\_BIAS\_256\_1 & 12 & 32 & 0.09375 & 0.1171875 & 96 \\ \hline
        DSAC\_256\_1 & 12 & 36 & 0.0625 & 0.1015625 & 92 \\ \hline
        DBIC\_256\_1 & 12 & 36 & 0.109375 & 0.0703125 & 92 \\ \hline
        NL\_256\_1 & 12 & 32 & 0.109375 & 0.109375 & 96 \\ \hline
        MEAN\_256\_1 & 11.34458 & 35.31399999 & 0.11402515 & 0.11379234 & 92.52599999 \\ \hline
        DU\_Rij\_Cyc & 8 & 34 & 0.125 & 0.1328125 & 94 \\ \hline
        MAX\_BIAS Rij\_Cyc & 12 & 30 & 0.109375 & 0.109375 & 98 \\ \hline
        DSAC\_Rij\_Cyc & 12 & 34 & 0.0625 & 0.1015625 & 94 \\ \hline
        DBIC\_Rij\_Cyc & 12 & 36 & 0.09375 & 0.0625 & 92 \\ \hline
        NL\_Rij\_Cyc & 10 & 30 & 0.09375 & 0.109375 & 98 \\ \hline
        MEAN\_Rij\_Cyc & 11.345 & 35.43999998 & 0.11402484 & 0.11383734 & 92.701 \\ \hline
    \end{tabular}
    \caption{Cryptographic Properties of Generated Sboxes}
    \label{tablemetrics}
\end{table}

\newpage

\newpage

\newpage

\newpage
\section{Experimental Cipher Setup}
In the second part of the experiment, we set up a simple SPN cipher, with fixed permutation layer and key schedule; and report the overall full cipher avalanche effect using four S-boxes: DBIC\_Rij\_Cyc, NL\_64\_4, DSAC\_Random and the AES S-box.

\subsection{Cipher Description}
The cipher takes a 64-bit, or 8-byte input plaintext and performs byte-level substitution using the examined 8x8 S-box, a byte-level permutation using the one-cycle permutation pbox\_8 $(2, 7, 1, 5, 0, 6, 4, 3)$ (0-indexed notation), a bit-level permutation using pbox\_64 (\ref{dillperm}-Dillon's permutation), then a XOR with the current round key. The key schedule, inspired from lightweight ciphers, is as follows: current key is formed from previous key, in the first round the previous key is the input master key of 8 bytes. Previous key is rotated left once, bytewise, then each nibble (group of 4 bits, or number from 0 to 15) is substituted according to a fixed key 4x4 S-box used in the PRESENT cipher: $[0, 6, 12, 1,5, 9, 11, 14, 3, 13, 15, 8, 10, 7, 4, 2]$. The first byte of the intermediate key is then XORed with the round constant (the round index $r$ in this case) and finally the result is XORed with the previous key bytewise rotated left three times. This generates the list of round keys. Below are the visual descriptions of the cipher and the key schedule.

\begin{figure}[h!]
    \centering
    \includegraphics[width=0.9\textwidth]{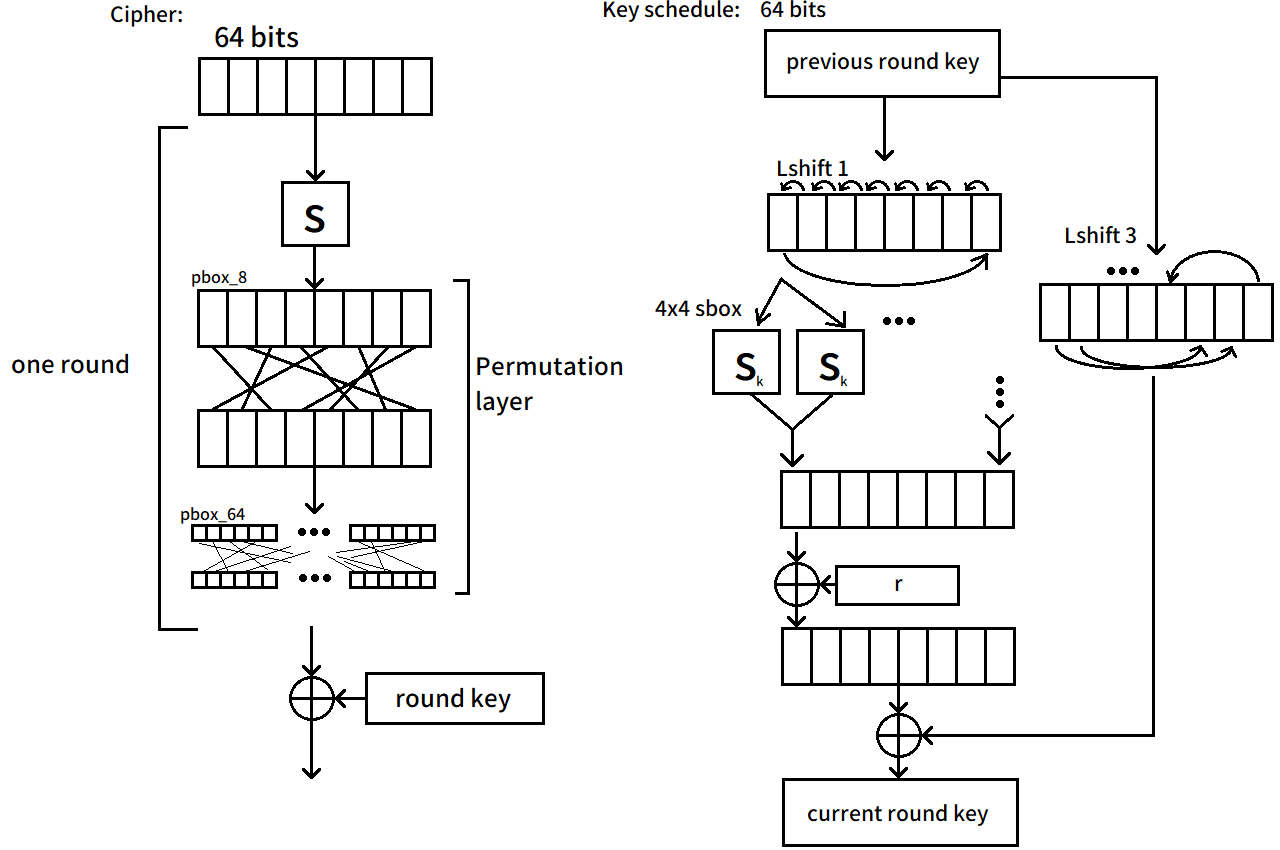}
    \caption{SPN cipher visual description}
    \label{ciph}
\end{figure}

\subsection{Per Cipher Avalanche}
To evaluate the avalanche effect at the cipher level, similar to the DSAC discussed before, we measure the change in ciphertexts resulting from single-bit flips in the plaintext. For each S-box under consideration, a large set of random plaintext–key pairs is generated, and for every trial a baseline ciphertext is computed. Each input bit is then individually flipped and the corresponding ciphertext is recalculated. The Hamming distance (number of bit-flips) between the baseline and perturbed ciphertexts is recorded and across all trials, yielding a distribution of avalanche values for each S-box. The resulting measures provide an empirical assessment of how closely each cipher approaches the ideal avalanche criterion, thereby enabling direct comparison between the proposed constructions and the AES standard. 

\subsection{Comparison of used S-boxes}
The experiment was done using the same set of randomly generated 10000 plaintext-key pairs for 4, 6 and 12 rounds of the cipher for all 4 S-boxes. The table \ref{ciphertab} represents the mean absolute distance from the perfect value - 32 of the 64 bits flipped, centered at 0. For example, using the AES S-box with 4 rounds yields, on average, \~32.82 or \~31.18 bit flips. Although we observe good (small) values for some isolated cases (DSAC\_Random 6 rounds) we have to consider that our 10000 key - cipher text pairs could be too small of a sample out of $2^{64}\times 2^{64}$ total pairs. Another important observation is that, as expected, the number of rounds in the cipher greatly improves the diffusion, which is apparent regardless of sample size. There is good space to improve this analysis, for example, using the AES linear layers and swapping the S-box only instead of our simple cipher, or considering more plaintext-ciphertext pairs (more computationally expensive) for more accurate results.

\begin{table}[!ht]
    \centering
    \begin{tabular}{|l|l|l|l|}
    \hline
        ~ & 4 rounds & 6 rounds & 12 rounds \\ \hline
        AES & 0.82310937 & 0.008609375 & 0.0016359375 \\ \hline
        DSAC\_Random & 0.89609374 & 0.006 & 0.0024375 \\ \hline
        DBIC\_Rij\_Cyc & 0.835890625 & 0.028421875 & 0.0165 \\ \hline
        NL\_64\_4 & 0.71510937 & 0.0055 & 0.00453125 \\ \hline
    \end{tabular}
    \caption{Distance from perfect diffusion}
    \label{ciphertab}
\end{table}

\chapter{Conclusion and Future Work}
\section{Insights for Future S-box Generation}
In this dissertation we have examined the main cryptographic properties that determine the strength of S-boxes, including differential uniformity, nonlinearity, avalanche, bit independence and properties of the ANF. Each property reveals potential vulnerabilities to differential, linear and algebraic attacks respectively. Mathematical constructions, such as those based on the inverse in AES or almost-bent constructions continue to offer strong theoretical guarantees and predictable strength, making them a more suitable approach. However, random generation remains a simpler and more flexible approach, better suited for smaller S-box sizes to be used in lightweight ciphers.

\section{Reflection and Learning}
Through this project I gained a deeper understanding of modern cryptanalysis, particularly about the design and evaluation of S-boxes and the studies on special classes of functions such as APN and Bent. Exploring the principles and layout of symmetric block ciphers gave me insight into the connection between theoretical concepts and final design choices. The project also strengthened my technical skills: I developed the analysis tools from scratch, without predefined libraries and functions, learned Julia to complement the readability of Python with high computational efficiency, and produced code fit for experimental use. I developed transferable research skills, such as finding and synthesising academic papers and writing in a structured, formal style. The project was not short of drawbacks, from choosing a topic to moving between theoretical and practical approaches and managing time around set milestones. These experiences not only exposed me to the current state of this branch of cryptography but also equipped me with skills to support further research in the field.

\section{Potential Research Directions}
Methods are constantly being developed with some being purely mathematical to more hybrid constructions. An example of the latter is the recent \cite{frob}, where authors, inspired by the design of AES which uses the inverse function, use the Frobenius endomorphism $f(x)=x^p$ (automorphism in finite fields), where $p$ is 2 as our fields have characteristic 2 ($1\oplus 1=0$) together with Möbius transformation in the finite group $f(x)=(ax\oplus b)\cdot(cx\oplus d)^{-1}$ where $a,b,c,d\in\mathbf{GF(2^8)}$ and the inverse is taken as before with respect to an irreducible. Another relevant example explained here leans more into the randomness side \cite{mordel}. Emphasis has been also put on constraining some properties of the S-box such as in \cite{hongjun}, where fixed and reverse fixed points are entirely removed before using an improved coupling quadratic map (a 3D system of linked chaos-like equations sensitive to initial value conditions) with an initial "seed" or "key" for reproducible results. Another similar example would be to enforce a number of Trace 0 elements to map to Trace 0 elements in the permutation. As such the experiment here can be greatly improved with more specific constraints, computational power and statistical analysis on the results. Generating good S-boxes remains an essential research topic; even as we move towards the future of quantum computers S-boxes remain an essential component which insures privacy at both wide and small scales.


\appendix
\chapter {S-boxes from the table \ref{tablemetrics}}
\label{appendix:sboxes}

S-boxes are presented as 0-indexed lists of 256 elements, with the meaning $S(i)=S[i]$ where $i=0, 1, 2, ...,255$ \newline
\begin{lstlisting}[caption={List of all S-boxes found}, language=, breaklines=true]
DU_Rij_Cyc = [52, 147, 132, 238, 103, 44, 14, 248, 175, 31, 12, 102, 20, 66, 135, 43, 19, 94, 246, 161, 181, 225, 251, 29, 222, 198, 155, 160, 8, 122, 191, 83, 193, 226, 206, 195, 162, 224, 142, 218, 202, 179, 3, 235, 34, 203, 124, 158, 201, 210, 37, 215, 151, 127, 183, 72, 45, 189, 230, 104, 231, 208, 216, 204, 233, 36, 4, 96, 16, 165, 212, 134, 163, 112, 239, 188, 18, 150, 207, 190, 186, 250, 63, 22, 105, 95, 64, 214, 49, 128, 126, 2, 247, 68, 28, 241, 77, 133, 125, 118, 89, 99, 123, 240, 48, 129, 110, 111, 38, 164, 136, 148, 32, 88, 166, 219, 217, 40, 84, 62, 138, 30, 87, 13, 54, 172, 90, 59, 73, 185, 27, 173, 145, 159, 245, 117, 75, 141, 109, 115, 174, 249, 211, 79, 41, 220, 200, 139, 177, 157, 255, 194, 
61, 60, 144, 243, 91, 130, 176, 187, 6, 86, 171, 74, 7, 227, 252, 57, 78, 167, 81, 192, 229, 58, 143, 205, 93, 17, 82, 47, 184, 170, 69, 153, 100, 106, 10, 101, 253, 152, 121, 76, 107, 223, 254, 137, 168, 71, 42, 80, 234, 51, 0, 178, 149, 196, 97, 56, 182, 67, 197, 70, 15, 244, 85, 33, 116, 113, 213, 221, 24, 55, 5, 1, 108, 146, 119, 209, 92, 169, 114, 242, 23, 21, 98, 46, 11, 120, 53, 50, 237, 180, 156, 39, 25, 9, 236, 232, 140, 228, 65, 154, 199, 35, 131, 26]

MAX_BIAS_Rij_Cyc = [203, 142, 104, 51, 85, 6, 131, 184, 50, 232, 158, 73, 110, 155, 33, 177, 212, 150, 182, 145, 252, 47, 2, 208, 125, 174, 70, 115, 62, 127, 226, 245, 71, 84, 21, 86, 218, 242, 228, 194, 205, 76, 128, 11, 193, 116, 32, 171, 66, 196, 34, 224, 12, 225, 90, 22, 0, 138, 152, 63, 243, 210, 8, 112, 209, 31, 45, 35, 100, 151, 114, 223, 143, 140, 141, 65, 57, 233, 176, 240, 183, 206, 108, 222, 123, 247, 250, 132, 124, 179, 13, 165, 229, 244, 101, 246, 175, 213, 156, 10, 97, 14, 167, 46, 211, 99, 26, 157, 148, 94, 15, 172, 121, 54, 230, 238, 195, 169, 191, 61, 154, 173, 49, 106, 178, 149, 227, 198, 9, 37, 207, 19, 72, 188, 41, 102, 20, 39, 162, 92, 3, 87, 200, 109, 130, 23, 117, 168, 251, 166, 68, 74, 137, 
103, 7, 28, 221, 5, 43, 234, 38, 147, 236, 80, 159, 190, 235, 1, 164, 197, 241, 77, 220, 88, 253, 67, 249, 231, 185, 105, 119, 255, 56, 248, 
139, 29, 216, 53, 59, 129, 96, 118, 180, 18, 160, 113, 98, 107, 187, 36, 204, 27, 122, 161, 199, 24, 42, 58, 237, 133, 201, 189, 186, 192, 181, 75, 89, 163, 30, 82, 135, 91, 215, 60, 120, 40, 17, 217, 81, 79, 254, 83, 136, 78, 170, 16, 64, 52, 214, 25, 134, 144, 95, 111, 153, 146, 
48, 44, 219, 93, 69, 202, 126, 55, 4, 239]

DSAC_Rij_Cyc = [214, 50, 54, 68, 120, 72, 232, 55, 209, 194, 200, 39, 30, 197, 91, 230, 37, 170, 78, 133, 79, 160, 84, 9, 183, 28, 96, 248, 18, 6, 188, 139, 207, 22, 61, 142, 85, 253, 210, 19, 34, 36, 23, 151, 254, 155, 92, 38, 60, 152, 204, 242, 70, 143, 239, 243, 74, 15, 165, 180, 122, 250, 110, 47, 109, 104, 140, 195, 83, 44, 176, 115, 71, 99, 205, 0, 221, 226, 2, 13, 148, 8, 89, 178, 48, 129, 218, 163, 215, 229, 186, 145, 146, 168, 105, 126, 187, 251, 167, 53, 193, 95, 29, 62, 35, 73, 224, 150, 26, 217, 27, 17, 228, 236, 124, 206, 1, 56, 88, 153, 141, 135, 65, 24, 128, 208, 130, 233, 33, 182, 219, 185, 63, 247, 11, 49, 154, 21, 87, 131, 171, 166, 106, 222, 127, 69, 64, 67, 100, 121, 5, 10, 3, 66, 213, 14, 201, 227, 157, 235, 12, 238, 169, 202, 114, 40, 223, 98, 81, 77, 4, 225, 181, 175, 41, 161, 246, 32, 134, 16, 252, 101, 198, 116, 220, 136, 112, 156, 123, 108, 144, 52, 237, 216, 113, 159, 59, 245, 211, 90, 20, 234, 25, 190, 57, 111, 192, 203, 244, 179, 103, 240, 174, 51, 184, 93, 118, 125, 46, 158, 80, 82, 45, 162, 43, 147, 31, 107, 102, 255, 199, 7, 212, 119, 173, 75, 97, 42, 137, 76, 191, 58, 196, 172, 117, 231, 241, 
249, 138, 94, 189, 149, 177, 86, 164, 132]

DBIC_Rij_Cyc = [188, 68, 119, 98, 48, 62, 159, 61, 34, 127, 76, 133, 221, 213, 105, 163, 214, 64, 59, 211, 231, 165, 198, 207, 123, 251, 131, 230, 177, 205, 180, 14, 145, 158, 218, 19, 88, 28, 139, 92, 199, 238, 93, 233, 247, 201, 35, 220, 246, 156, 226, 73, 175, 107, 152, 171, 69, 140, 194, 99, 
186, 103, 189, 243, 147, 45, 210, 237, 30, 106, 240, 254, 31, 202, 75, 118, 67, 109, 190, 185, 49, 164, 102, 24, 228, 83, 146, 204, 157, 244, 174, 55, 255, 223, 142, 3, 197, 120, 101, 135, 160, 82, 241, 125, 15, 11, 132, 130, 63, 78, 173, 40, 232, 70, 225, 47, 33, 209, 161, 6, 13, 
229, 91, 245, 100, 235, 250, 12, 137, 162, 77, 169, 179, 196, 53, 36, 200, 155, 192, 10, 57, 17, 116, 23, 96, 183, 168, 80, 114, 97, 112, 154, 41, 121, 8, 90, 22, 144, 84, 60, 143, 52, 242, 111, 178, 1, 206, 193, 222, 29, 208, 26, 74, 227, 219, 217, 252, 216, 56, 249, 46, 191, 39, 
66, 253, 167, 25, 85, 115, 37, 166, 50, 248, 122, 187, 117, 236, 87, 72, 54, 0, 150, 4, 234, 42, 86, 151, 181, 38, 79, 20, 44, 95, 129, 81, 203, 71, 176, 212, 182, 89, 184, 215, 5, 16, 104, 9, 153, 32, 148, 224, 18, 136, 27, 239, 126, 124, 65, 134, 113, 195, 128, 170, 21, 94, 108, 
110, 51, 149, 172, 138, 7, 141, 43, 2, 58]

NL_Rij_Cyc = [138, 51, 181, 52, 190, 246, 227, 113, 153, 11, 225, 1, 241, 164, 211, 50, 116, 16, 83, 180, 206, 22, 133, 158, 193, 62, 111, 141, 4, 86, 115, 212, 88, 230, 143, 66, 118, 107, 114, 74, 155, 25, 124, 9, 91, 201, 215, 99, 250, 24, 33, 251, 253, 248, 39, 217, 136, 120, 184, 200, 234, 
35, 254, 58, 239, 27, 81, 159, 34, 142, 131, 112, 89, 28, 84, 64, 228, 108, 209, 47, 126, 63, 109, 106, 242, 218, 45, 182, 185, 80, 199, 198, 17, 177, 23, 197, 235, 125, 105, 151, 244, 166, 232, 243, 21, 79, 189, 85, 238, 146, 226, 154, 208, 240, 76, 150, 121, 6, 31, 188, 44, 140, 
196, 175, 219, 204, 135, 53, 94, 147, 87, 98, 245, 61, 3, 119, 176, 202, 167, 194, 221, 102, 103, 165, 127, 2, 129, 5, 160, 191, 38, 72, 104, 42, 41, 249, 179, 144, 157, 237, 117, 183, 56, 186, 252, 223, 163, 210, 231, 216, 0, 123, 78, 156, 229, 57, 170, 43, 46, 97, 20, 10, 236, 132, 96, 187, 173, 205, 18, 134, 137, 13, 92, 222, 161, 207, 149, 178, 100, 7, 59, 214, 171, 54, 90, 95, 122, 75, 26, 255, 110, 65, 174, 49, 69, 128, 220, 77, 73, 162, 60, 71, 247, 8, 148, 93, 67, 168, 37, 139, 36, 172, 55, 169, 14, 203, 68, 152, 82, 145, 32, 19, 48, 213, 30, 224, 101, 195, 15, 29, 12, 70, 192, 40, 130, 233]

DU_256_1 = [182, 46, 139, 33, 8, 247, 29, 207, 134, 78, 248, 210, 215, 218, 23, 97, 235, 5, 226, 164, 107, 141, 128, 250, 35, 200, 79, 206, 245, 143, 138, 43, 174, 2, 149, 31, 137, 196, 119, 1, 190, 19, 178, 148, 153, 133, 49, 219, 65, 152, 201, 232, 175, 202, 85, 176, 185, 217, 64, 216, 69, 
55, 160, 42, 225, 95, 194, 66, 109, 249, 115, 156, 172, 163, 241, 45, 90, 73, 171, 3, 34, 130, 145, 86, 154, 131, 24, 9, 47, 92, 7, 36, 244, 
222, 181, 242, 106, 60, 100, 208, 193, 54, 17, 214, 13, 111, 62, 135, 188, 151, 99, 37, 129, 227, 203, 162, 142, 70, 146, 80, 144, 59, 104, 20, 220, 81, 189, 56, 27, 83, 18, 161, 67, 121, 136, 140, 122, 197, 205, 168, 12, 123, 10, 179, 191, 71, 14, 184, 253, 101, 26, 116, 252, 126, 132, 186, 84, 127, 195, 118, 68, 50, 32, 150, 169, 38, 58, 103, 224, 213, 63, 39, 204, 11, 159, 0, 93, 239, 72, 22, 155, 98, 114, 238, 117, 
212, 166, 112, 89, 236, 61, 124, 198, 52, 237, 243, 76, 170, 221, 254, 113, 110, 87, 21, 125, 211, 240, 234, 228, 246, 233, 57, 177, 74, 120, 167, 96, 94, 192, 4, 183, 199, 223, 251, 53, 147, 51, 108, 229, 28, 82, 48, 15, 180, 255, 173, 102, 6, 187, 30, 77, 40, 88, 165, 44, 157, 16, 231, 158, 230, 91, 25, 41, 209, 105, 75]

MAX_BIAS_256_1 = [230, 160, 43, 48, 189, 118, 40, 116, 39, 31, 169, 249, 49, 140, 18, 133, 131, 60, 111, 70, 102, 95, 238, 157, 129, 135, 79, 91, 245, 68, 109, 4, 190, 187, 62, 52, 145, 59, 55, 35, 206, 159, 244, 154, 195, 227, 196, 13, 193, 166, 2, 125, 90, 191, 32, 177, 205, 105, 44, 15, 235, 141, 181, 137, 208, 47, 10, 94, 128, 217, 158, 51, 224, 97, 92, 246, 65, 199, 237, 72, 123, 143, 194, 20, 241, 231, 7, 112, 1, 226, 17, 252, 254, 147, 204, 71, 182, 223, 50, 178, 229, 66, 179, 77, 16, 64, 149, 171, 202, 176, 207, 6, 167, 110, 148, 98, 212, 218, 53, 45, 172, 220, 33, 23, 200, 136, 239, 240, 213, 69, 174, 42, 115, 168, 41, 216, 86, 144, 180, 82, 26, 175, 25, 38, 108, 153, 28, 87, 242, 56, 46, 234, 161, 9, 96, 3, 228, 250, 155, 236, 165, 104, 8, 54, 126, 232, 100, 121, 219, 93, 162, 85, 138, 192, 37, 201, 150, 134, 11, 89, 30, 222, 170, 57, 215, 27, 81, 21, 88, 197, 255, 184, 124, 209, 0, 101, 119, 74, 225, 36, 203, 12, 152, 122, 83, 76, 188, 19, 99, 67, 22, 63, 5, 210, 163, 61, 214, 139, 114, 24, 156, 14, 117, 164, 253, 29, 185, 211, 151, 113, 248, 142, 233, 120, 221, 84, 243, 80, 130, 146, 106, 58, 73, 132, 75, 247, 34, 107, 78, 103, 173, 198, 183, 251, 186, 127]

DSAC_256_1 = [46, 254, 201, 185, 207, 67, 220, 25, 13, 134, 70, 139, 1, 76, 149, 72, 158, 87, 227, 121, 163, 82, 195, 113, 181, 214, 90, 142, 135, 77, 31, 71, 194, 125, 237, 122, 187, 235, 130, 123, 133, 108, 86, 174, 124, 146, 150, 180, 50, 190, 204, 99, 38, 49, 40, 53, 157, 191, 118, 211, 200, 110, 230, 166, 5, 45, 127, 47, 24, 119, 208, 137, 255, 244, 129, 161, 73, 233, 198, 56, 236, 51, 6, 41, 151, 98, 58, 131, 205, 44, 91, 138, 177, 101, 196, 75, 29, 42, 84, 26, 243, 172, 192, 182, 20, 17, 28, 88, 242, 169, 253, 78, 12, 120, 245, 231, 128, 224, 203, 54, 183, 148, 43, 170, 206, 65, 74, 95, 61, 0, 16, 247, 4, 189, 164, 248, 222, 100, 188, 55, 104, 59, 252, 102, 106, 15, 250, 117, 228, 147, 89, 154, 62, 79, 212, 94, 223, 221, 22, 96, 109, 63, 171, 32, 19, 105, 85, 159, 66, 48, 153, 213, 140, 210, 238, 232, 60, 103, 155, 178, 34, 3, 18, 52, 241, 
202, 9, 197, 193, 143, 234, 8, 186, 136, 156, 176, 83, 217, 81, 226, 239, 225, 251, 184, 144, 218, 165, 145, 116, 132, 2, 111, 30, 64, 36, 160, 112, 97, 126, 35, 215, 167, 92, 115, 114, 80, 14, 175, 219, 93, 141, 23, 10, 7, 162, 11, 168, 173, 57, 39, 229, 37, 107, 216, 68, 240, 69, 209, 199, 27, 21, 179, 152, 33, 246, 249]

DBIC_256_1 = [149, 54, 40, 67, 117, 183, 39, 83, 177, 145, 234, 179, 99, 209, 8, 123, 233, 138, 49, 68, 199, 204, 174, 212, 236, 171, 134, 133, 114, 125, 
36, 10, 206, 46, 146, 57, 147, 127, 55, 158, 144, 191, 43, 143, 131, 6, 79, 41, 96, 172, 157, 113, 160, 186, 20, 71, 202, 247, 45, 194, 213,
87, 196, 104, 72, 122, 11, 156, 244, 192, 84, 48, 1, 15, 103, 65, 222, 164, 167, 16, 111, 69, 170, 115, 105, 124, 42, 53, 75, 220, 218, 129, 
28, 162, 132, 29, 231, 224, 76, 107, 94, 58, 232, 60, 3, 118, 243, 25, 185, 221, 189, 135, 241, 237, 159, 30, 136, 163, 2, 255, 228, 50, 251, 14, 219, 207, 254, 93, 98, 190, 246, 106, 17, 21, 19, 100, 197, 78, 37, 210, 0, 81, 205, 73, 240, 44, 23, 126, 80, 9, 61, 211, 249, 26, 151, 88, 130, 110, 155, 223, 51, 229, 139, 64, 154, 86, 152, 245, 142, 230, 200, 4, 33, 34, 141, 188, 198, 47, 203, 161, 85, 52, 24, 77, 168, 176, 180, 248, 208, 5, 214, 215, 91, 59, 119, 101, 238, 112, 32, 70, 120, 74, 140, 66, 181, 165, 89, 18, 195, 252, 242, 31, 116, 56, 173, 250, 35, 216, 239, 253, 38, 184, 169, 128, 166, 22, 108, 201, 226, 137, 12, 97, 7, 148, 27, 178, 92, 225, 235, 217, 82, 63, 187, 90, 227, 150, 95, 
121, 13, 175, 153, 109, 193, 102, 182, 62]

NL_256_1 = [149, 174, 214, 46, 22, 135, 67, 200, 253, 120, 114, 65, 10, 100, 51, 210, 91, 94, 183, 111, 250, 187, 170, 98, 134, 177, 243, 196, 14, 77, 83, 23, 57, 93, 215, 24, 173, 184, 255, 208, 189, 63, 29, 45, 182, 236, 199, 104, 72, 137, 181, 226, 251, 32, 190, 18, 222, 59, 62, 178, 50, 224, 218, 155, 47, 223, 41, 213, 153, 84, 132, 70, 49, 52, 78, 0, 234, 216, 175, 169, 3, 115, 254, 133, 192, 88, 247, 4, 238, 206, 233, 172, 9, 168, 124, 75, 112, 229, 121, 53, 126, 235, 167, 217, 139, 107, 163, 54, 230, 252, 194, 143, 157, 154, 148, 7, 191, 164, 82, 158, 227, 95, 81, 61, 123, 188, 201, 131, 142, 73, 40, 31, 165, 92, 48, 239, 64, 240, 25, 103, 34, 86, 198, 27, 105, 185, 127, 8, 96, 108, 118, 1, 231, 159, 156, 232, 140, 150, 102, 13, 219, 186, 145, 26, 193, 97, 58, 225, 130, 244, 109, 125, 87, 74, 249, 71, 15, 101, 162, 56, 171, 248, 89, 43, 55, 35, 203, 117, 113, 241, 212, 141, 44, 90, 19, 176, 37, 128, 2, 202, 228, 42, 129, 116, 85, 39, 33, 38, 246, 166, 146, 144, 221, 205, 16, 119, 209, 242, 6, 151, 180, 237, 99, 30, 5, 161, 20, 36, 138, 60, 28, 106, 136, 160, 122, 11, 66, 110, 12, 245, 80, 68, 17, 69, 220, 179, 211, 152, 21, 147, 204, 195, 79, 207, 76, 197]

DU_64_4 = [180, 35, 174, 135, 251, 125, 132, 72, 107, 10, 173, 106, 14, 30, 255, 65, 23, 29, 229, 162, 74, 222, 129, 254, 226, 144, 50, 83, 43, 181, 119, 73, 240, 52, 54, 95, 78, 111, 11, 112, 34, 149, 12, 190, 252, 239, 24, 147, 90, 244, 150, 172, 81, 126, 94, 115, 131, 155, 189, 145, 236, 
217, 67, 127, 71, 157, 2, 104, 17, 105, 198, 214, 169, 233, 116, 158, 20, 192, 215, 75, 146, 114, 160, 139, 55, 156, 199, 13, 4, 58, 166, 207, 46, 249, 88, 243, 133, 80, 9, 175, 225, 89, 33, 140, 53, 56, 113, 84, 64, 230, 194, 163, 19, 196, 118, 186, 191, 148, 232, 188, 57, 69, 224, 208, 92, 36, 45, 96, 31, 32, 200, 7, 77, 124, 120, 44, 130, 165, 152, 93, 63, 237, 97, 248, 206, 16, 108, 220, 66, 250, 76, 8, 128, 3, 38, 
154, 209, 85, 15, 234, 27, 102, 201, 197, 182, 247, 210, 100, 178, 60, 70, 51, 176, 227, 136, 82, 195, 121, 211, 22, 202, 253, 231, 87, 168, 
6, 184, 123, 235, 213, 171, 137, 179, 159, 47, 228, 193, 242, 143, 241, 109, 110, 5, 161, 62, 138, 142, 117, 91, 218, 98, 40, 1, 37, 216, 28, 151, 42, 141, 59, 245, 41, 26, 219, 49, 153, 79, 164, 103, 48, 68, 183, 39, 204, 101, 221, 167, 177, 25, 134, 86, 21, 185, 0, 99, 61, 170, 205, 122, 18, 187, 203, 212, 246, 238, 223]

MAX_BIAS_64_4 = [42, 4, 246, 137, 241, 94, 200, 189, 120, 242, 251, 149, 255, 74, 90, 209, 193, 223, 244, 237, 236, 150, 59, 157, 122, 65, 48, 220, 153, 214, 25, 228, 202, 29, 115, 219, 159, 227, 231, 83, 69, 126, 6, 20, 213, 191, 43, 110, 82, 158, 72, 254, 99, 86, 22, 32, 51, 93, 11, 171, 39, 212, 81, 68, 79, 127, 170, 129, 76, 238, 35, 118, 134, 151, 204, 67, 57, 172, 208, 45, 16, 101, 195, 107, 106, 61, 77, 132, 176, 143, 161, 116, 
218, 196, 85, 50, 169, 141, 7, 26, 125, 140, 66, 210, 243, 249, 142, 144, 167, 186, 49, 197, 201, 180, 130, 179, 10, 91, 224, 162, 229, 46, 71, 53, 84, 97, 96, 54, 177, 78, 27, 0, 103, 234, 40, 114, 133, 14, 248, 63, 145, 62, 199, 226, 175, 5, 92, 124, 185, 245, 123, 155, 131, 12, 
183, 146, 105, 24, 17, 112, 198, 205, 182, 168, 80, 166, 30, 207, 192, 18, 1, 138, 139, 163, 165, 102, 31, 188, 240, 203, 3, 64, 98, 33, 89, 
174, 222, 253, 52, 152, 187, 88, 135, 36, 47, 239, 194, 34, 184, 232, 235, 87, 211, 100, 154, 37, 2, 121, 44, 41, 178, 117, 225, 156, 104, 108, 19, 109, 173, 75, 136, 164, 8, 119, 181, 217, 60, 230, 250, 148, 111, 95, 13, 221, 160, 9, 56, 15, 28, 58, 215, 233, 216, 252, 23, 70, 128, 55, 247, 147, 21, 113, 38, 206, 73, 190]

DSAC_64_4 = [189, 45, 43, 187, 0, 253, 76, 215, 232, 110, 153, 150, 222, 30, 66, 254, 251, 50, 131, 242, 80, 183, 204, 77, 90, 211, 88, 171, 91, 147, 235, 177, 198, 34, 9, 142, 145, 197, 185, 227, 228, 146, 218, 32, 4, 172, 65, 40, 136, 38, 107, 135, 125, 126, 86, 109, 194, 186, 205, 224, 243, 175, 57, 44, 102, 35, 101, 1, 118, 134, 103, 129, 179, 229, 36, 248, 111, 188, 121, 234, 120, 244, 203, 178, 28, 143, 116, 139, 167, 217, 59, 6, 246, 128, 61, 119, 169, 149, 68, 100, 168, 26, 63, 255, 49, 191, 250, 237, 238, 92, 108, 239, 27, 122, 41, 7, 157, 181, 182, 67, 208, 93, 5, 95, 46, 196, 133, 247, 202, 221, 214, 19, 166, 174, 114, 16, 75, 11, 207, 84, 249, 195, 82, 192, 14, 130, 83, 124, 51, 225, 230, 15, 55, 31, 158, 106, 62, 210, 141, 151, 138, 115, 22, 190, 24, 20, 64, 127, 201, 159, 79, 56, 240, 54, 216, 23, 81, 78, 97, 220, 73, 173, 39, 252, 
112, 165, 37, 18, 89, 69, 17, 245, 98, 231, 212, 21, 206, 47, 213, 226, 25, 3, 74, 85, 117, 155, 13, 180, 148, 12, 154, 164, 71, 144, 162, 72, 96, 137, 209, 199, 8, 233, 236, 160, 193, 140, 123, 10, 113, 99, 200, 161, 176, 29, 42, 104, 58, 94, 53, 33, 163, 152, 2, 52, 156, 132, 223, 60, 241, 87, 105, 48, 70, 219, 170, 184]

DBIC_64_4 = [239, 245, 204, 37, 162, 91, 10, 180, 123, 69, 148, 7, 128, 179, 113, 135, 160, 155, 129, 44, 62, 3, 255, 141, 12, 94, 159, 216, 240, 243, 158, 196, 67, 60, 169, 227, 106, 110, 207, 117, 208, 81, 124, 125, 51, 40, 222, 233, 143, 254, 54, 72, 34, 118, 38, 95, 68, 186, 78, 224, 193, 
2, 134, 236, 33, 102, 166, 86, 1, 157, 212, 189, 55, 215, 251, 115, 80, 176, 153, 178, 21, 46, 192, 213, 101, 19, 24, 88, 252, 79, 111, 173, 
42, 73, 149, 100, 104, 66, 201, 65, 8, 116, 58, 203, 11, 182, 136, 174, 187, 150, 43, 22, 109, 77, 163, 90, 49, 156, 30, 244, 83, 99, 197, 70, 98, 53, 133, 96, 0, 181, 41, 120, 27, 142, 76, 25, 228, 93, 64, 14, 246, 185, 167, 13, 31, 144, 171, 114, 29, 52, 20, 253, 161, 151, 195, 206, 18, 35, 28, 219, 85, 121, 107, 103, 190, 122, 132, 50, 61, 230, 57, 249, 247, 36, 138, 108, 16, 26, 172, 210, 5, 198, 237, 56, 229, 9, 39, 17, 235, 127, 165, 232, 92, 47, 97, 191, 112, 32, 242, 152, 154, 59, 45, 248, 231, 74, 130, 205, 200, 194, 214, 71, 164, 87, 199, 48, 105, 
184, 217, 126, 223, 140, 137, 89, 6, 75, 202, 82, 250, 170, 146, 23, 238, 139, 84, 119, 177, 234, 220, 63, 131, 175, 168, 209, 4, 218, 147, 241, 145, 226, 225, 188, 211, 15, 183, 221]

NL_64_4 = [57, 187, 25, 108, 53, 65, 118, 2, 101, 232, 152, 190, 225, 217, 255, 230, 84, 110, 72, 131, 224, 10, 106, 62, 31, 1, 107, 207, 3, 76, 138, 159, 199, 121, 114, 15, 158, 13, 148, 151, 126, 86, 164, 147, 137, 128, 60, 136, 181, 247, 34, 209, 71, 205, 218, 191, 30, 167, 172, 229, 161, 132, 192, 32, 88, 165, 141, 55, 73, 245, 35, 41, 194, 105, 169, 153, 24, 142, 175, 238, 97, 75, 202, 130, 119, 16, 198, 177, 180, 91, 17, 185, 113, 250, 28, 227, 236, 95, 81, 54, 160, 156, 182, 102, 195, 21, 176, 155, 103, 116, 6, 210, 70, 221, 52, 83, 23, 183, 51, 63, 251, 123, 163, 22, 237, 94, 193, 19, 78, 211, 20, 206, 239, 9, 125, 48, 14, 248, 29, 18, 228, 124, 93, 212, 11, 8, 134, 117, 82, 171, 143, 214, 231, 47, 216, 144, 49, 61, 162, 223, 140, 87, 189, 249, 201, 43, 243, 36, 208, 129, 59, 254, 233, 203, 235, 196, 179, 188, 242, 0, 112, 80, 178, 220, 104, 33, 197, 39, 7, 184, 219, 66, 170, 67, 146, 244, 135, 226, 157, 64, 166, 89, 253, 77, 150, 56, 4, 168, 149, 58, 174, 133, 44, 69, 145, 
122, 213, 241, 154, 234, 98, 173, 186, 96, 26, 46, 37, 38, 92, 5, 90, 139, 85, 27, 99, 215, 115, 109, 200, 111, 127, 12, 204, 74, 240, 42, 68, 40, 45, 79, 252, 246, 120, 100, 50, 222]

DU_16_16 = [84, 178, 60, 63, 155, 142, 206, 77, 7, 94, 120, 239, 109, 188, 19, 236, 23, 133, 47, 93, 243, 218, 190, 61, 15, 9, 50, 70, 35, 166, 57, 122, 52, 37, 26, 117, 3, 162, 241, 5, 16, 21, 217, 237, 64, 4, 121, 42, 130, 184, 189, 114, 34, 95, 202, 27, 229, 156, 53, 227, 138, 192, 14, 251, 55, 193, 147, 76, 194, 253, 48, 86, 78, 8, 97, 69, 230, 51, 6, 219, 223, 161, 177, 46, 54, 176, 113, 2, 72, 44, 225, 150, 127, 214, 191, 210, 242, 228, 199, 112, 244, 92, 67, 144, 22, 17, 170, 30, 115, 106, 28, 25, 240, 185, 179, 212, 119, 163, 45, 172, 40, 59, 209, 158, 56, 73, 
104, 118, 222, 220, 207, 58, 90, 85, 197, 187, 175, 32, 171, 216, 238, 1, 131, 105, 183, 24, 71, 126, 201, 116, 154, 111, 182, 196, 89, 125, 
38, 145, 141, 180, 29, 165, 164, 254, 74, 157, 96, 231, 49, 101, 135, 248, 160, 153, 255, 167, 173, 149, 20, 169, 232, 132, 102, 82, 203, 75, 252, 205, 137, 247, 181, 41, 159, 11, 233, 79, 12, 148, 128, 91, 68, 10, 215, 107, 221, 143, 152, 43, 200, 224, 81, 226, 146, 168, 66, 139, 
99, 33, 235, 129, 87, 18, 174, 103, 213, 245, 136, 195, 250, 108, 36, 13, 100, 123, 151, 198, 39, 98, 124, 208, 110, 246, 80, 65, 134, 62, 31, 211, 83, 88, 186, 249, 204, 140, 0, 234]

MAX_BIAS_16_16 = [81, 217, 9, 142, 104, 234, 91, 248, 130, 223, 191, 45, 197, 5, 204, 158, 74, 100, 140, 57, 151, 189, 188, 126, 96, 135, 201, 72, 108, 8, 42, 192, 160, 95, 128, 252, 44, 38, 168, 68, 58, 157, 103, 120, 186, 255, 193, 147, 184, 4, 155, 106, 15, 61, 35, 131, 227, 43, 125, 161, 54, 156, 32, 190, 224, 34, 7, 148, 136, 250, 180, 39, 203, 246, 112, 59, 22, 194, 66, 121, 75, 26, 165, 187, 19, 183, 173, 33, 176, 220, 13, 154, 181, 102, 29, 90, 2, 132, 134, 85, 226, 219, 14, 254, 62, 171, 251, 214, 239, 0, 60, 146, 243, 123, 76, 144, 82, 212, 221, 162, 20, 119, 170, 
49, 84, 196, 69, 178, 247, 174, 249, 141, 172, 167, 23, 225, 73, 50, 12, 253, 37, 169, 113, 70, 216, 110, 71, 229, 207, 24, 64, 31, 16, 48, 205, 10, 1, 129, 230, 27, 233, 199, 77, 25, 208, 232, 244, 99, 88, 28, 18, 79, 36, 198, 213, 115, 210, 83, 86, 137, 30, 206, 200, 67, 118, 237, 149, 52, 80, 107, 17, 215, 222, 179, 6, 241, 56, 47, 51, 89, 211, 231, 238, 93, 78, 175, 97, 55, 101, 117, 236, 218, 116, 3, 182, 185, 105, 114, 143, 65, 53, 122, 152, 228, 240, 98, 195, 163, 92, 150, 209, 138, 139, 41, 202, 11, 153, 94, 63, 242, 109, 87, 133, 124, 111, 21, 145, 
127, 235, 46, 40, 164, 166, 177, 245, 159]

DSAC_16_16 = [134, 12, 19, 214, 255, 142, 143, 235, 221, 110, 14, 215, 5, 231, 180, 121, 77, 198, 174, 133, 34, 186, 166, 104, 94, 66, 0, 205, 209, 156, 167, 254, 163, 48, 204, 234, 58, 249, 135, 211, 37, 67, 119, 126, 229, 169, 250, 193, 129, 105, 228, 22, 107, 176, 95, 148, 123, 237, 210, 53, 138, 161, 83, 251, 92, 197, 1, 43, 155, 196, 21, 59, 220, 131, 185, 8, 10, 217, 187, 190, 99, 73, 24, 172, 173, 75, 91, 222, 122, 239, 201, 
246, 27, 100, 36, 238, 151, 152, 125, 57, 7, 181, 2, 38, 158, 13, 51, 206, 127, 84, 15, 56, 194, 157, 236, 192, 165, 145, 112, 96, 20, 252, 195, 61, 3, 114, 90, 109, 149, 213, 164, 74, 62, 253, 116, 23, 35, 202, 86, 82, 117, 118, 49, 199, 224, 52, 39, 81, 101, 153, 93, 113, 88, 42, 242, 78, 170, 175, 144, 26, 64, 191, 87, 106, 232, 132, 223, 54, 44, 6, 200, 218, 136, 60, 241, 120, 216, 233, 32, 150, 178, 76, 160, 103, 31, 212, 25, 45, 30, 226, 171, 140, 247, 179, 207, 124, 17, 128, 46, 147, 245, 240, 89, 18, 65, 146, 85, 139, 137, 189, 244, 9, 68, 79, 162, 230, 243, 184, 47, 63, 208, 111, 227, 55, 177, 130, 16, 203, 80, 70, 141, 168, 248, 29, 115, 33, 225, 108, 72, 40, 219, 97, 102, 41, 11, 183, 
50, 159, 154, 188, 28, 71, 4, 98, 69, 182]

DBIC_16_16 = [41, 245, 171, 230, 198, 192, 72, 74, 123, 62, 66, 128, 3, 21, 223, 26, 235, 83, 34, 251, 233, 167, 172, 11, 109, 118, 150, 59, 131, 95, 38, 
100, 212, 102, 208, 7, 86, 183, 17, 187, 129, 56, 176, 121, 173, 252, 96, 154, 122, 20, 164, 87, 65, 103, 1, 54, 57, 28, 139, 213, 36, 90, 130, 80, 126, 179, 60, 149, 43, 155, 125, 228, 117, 248, 101, 30, 140, 166, 15, 53, 55, 159, 49, 188, 147, 162, 161, 25, 231, 44, 215, 75, 5, 199, 165, 151, 23, 16, 104, 77, 133, 175, 237, 157, 112, 219, 209, 88, 105, 197, 177, 239, 170, 195, 63, 12, 119, 221, 142, 160, 152, 148, 191, 2, 93, 211, 114, 194, 40, 94, 13, 68, 146, 204, 31, 84, 51, 205, 247, 37, 242, 81, 186, 236, 0, 78, 206, 32, 92, 180, 178, 127, 226, 214, 48, 39, 196, 113, 249, 138, 82, 9, 97, 190, 244, 217, 238, 115, 255, 91, 110, 108, 246, 229, 168, 135, 210, 24, 216, 67, 35, 61, 89, 116, 225, 6, 64, 243, 22, 42, 218, 169, 153, 174, 14, 33, 250, 145, 222, 29, 70, 46, 8, 200, 254, 76, 107, 182, 203, 19, 220, 232, 52, 201, 144, 4, 98, 45, 58, 143, 71, 50, 99, 106, 241, 120, 240, 141, 79, 202, 10, 253, 132, 111, 69, 156, 207, 189, 137, 163, 193, 85, 181, 184, 227, 136, 47, 158, 124, 185, 134, 73, 27, 18, 224, 234]

NL_16_16 = [22, 41, 249, 232, 255, 0, 148, 192, 117, 177, 78, 156, 243, 128, 39, 158, 92, 94, 131, 241, 210, 126, 54, 209, 150, 15, 55, 152, 239, 200, 67, 213, 173, 95, 225, 43, 27, 108, 48, 73, 160, 171, 53, 77, 197, 135, 161, 107, 151, 139, 216, 124, 159, 84, 32, 234, 98, 214, 113, 215, 62, 40, 186, 212, 34, 251, 233, 168, 123, 166, 89, 105, 187, 185, 91, 36, 228, 47, 203, 1, 165, 189, 246, 88, 80, 112, 115, 155, 202, 116, 60, 180, 188, 252, 83, 142, 104, 82, 74, 206, 182, 31, 230, 163, 2, 76, 208, 195, 181, 97, 70, 223, 242, 190, 69, 224, 86, 13, 248, 247, 147, 130, 138, 25, 9, 56, 65, 6, 176, 71, 30, 85, 58, 129, 46, 96, 167, 198, 81, 211, 50, 220, 122, 179, 201, 7, 193, 238, 141, 125, 101, 244, 144, 254, 103, 35, 38, 18, 20, 3, 162, 174, 184, 217, 205, 99, 231, 253, 140, 121, 33, 134, 11, 120, 111, 72, 172, 237, 154, 137, 240, 23, 207, 52, 
146, 61, 12, 59, 221, 16, 100, 222, 245, 183, 4, 63, 68, 204, 118, 93, 219, 178, 5, 109, 143, 229, 114, 29, 164, 133, 227, 218, 102, 75, 42, 
51, 110, 24, 87, 90, 196, 236, 106, 28, 169, 175, 10, 119, 66, 44, 49, 19, 14, 199, 157, 57, 136, 194, 153, 145, 226, 235, 45, 132, 21, 79, 149, 127, 191, 250, 26, 8, 37, 170, 17, 64]

DU_4_64 = [24, 240, 230, 182, 78, 28, 232, 99, 209, 70, 139, 210, 71, 57, 54, 165, 205, 222, 102, 190, 221, 167, 83, 254, 76, 123, 203, 236, 86, 1, 213, 14, 161, 199, 38, 198, 96, 148, 143, 193, 88, 160, 37, 64, 248, 212, 231, 90, 173, 62, 194, 40, 220, 183, 152, 49, 42, 15, 180, 98, 23, 243, 67, 80, 176, 106, 51, 55, 12, 197, 214, 151, 164, 89, 228, 84, 154, 85, 20, 116, 137, 166, 58, 26, 132, 156, 128, 16, 66, 35, 239, 79, 95, 
17, 200, 131, 50, 109, 44, 186, 147, 195, 81, 21, 216, 34, 48, 196, 224, 135, 30, 235, 142, 91, 191, 111, 113, 215, 126, 250, 7, 178, 8, 127, 170, 237, 204, 211, 5, 189, 253, 201, 225, 218, 136, 145, 177, 157, 134, 226, 155, 118, 101, 105, 103, 97, 77, 107, 56, 179, 227, 68, 31, 252, 0, 3, 146, 63, 130, 9, 104, 168, 29, 223, 233, 13, 18, 144, 169, 32, 174, 244, 124, 65, 172, 184, 187, 138, 33, 6, 192, 175, 140, 45, 207, 202, 120, 43, 246, 69, 117, 188, 82, 185, 36, 112, 100, 129, 73, 121, 74, 92, 39, 22, 141, 242, 115, 181, 27, 234, 119, 25, 53, 247, 159, 19, 41, 60, 61, 47, 171, 4, 255, 229, 150, 75, 249, 108, 94, 238, 46, 2, 149, 241, 122, 206, 245, 153, 163, 219, 162, 72, 87, 133, 52, 208, 114, 110, 59, 10, 11, 158, 125, 251, 217, 93]

MAX_BIAS_4_64 = [104, 174, 29, 87, 163, 126, 194, 120, 186, 16, 37, 24, 18, 20, 203, 192, 176, 229, 30, 12, 190, 69, 1, 196, 221, 175, 127, 162, 107, 90, 19, 117, 99, 151, 129, 179, 41, 157, 96, 198, 136, 154, 88, 247, 113, 153, 51, 180, 188, 156, 10, 172, 252, 114, 48, 199, 58, 52, 249, 38, 160, 
191, 67, 137, 177, 218, 134, 35, 14, 149, 28, 178, 140, 233, 54, 6, 216, 2, 44, 36, 132, 22, 49, 131, 25, 97, 144, 189, 195, 83, 77, 222, 45, 84, 72, 73, 161, 152, 53, 241, 234, 244, 209, 164, 236, 246, 94, 158, 167, 168, 159, 205, 15, 182, 171, 142, 213, 105, 0, 235, 184, 145, 111, 13, 71, 220, 212, 100, 243, 119, 95, 135, 55, 254, 138, 89, 215, 128, 197, 185, 106, 9, 183, 210, 102, 251, 60, 21, 112, 147, 201, 125, 253, 242, 79, 193, 217, 50, 70, 76, 231, 59, 39, 103, 4, 91, 5, 223, 187, 165, 124, 98, 250, 115, 81, 93, 141, 248, 170, 62, 61, 56, 78, 173, 227, 17, 155, 214, 74, 228, 123, 47, 148, 8, 230, 239, 240, 66, 27, 80, 121, 43, 204, 255, 116, 211, 23, 64, 101, 86, 133, 122, 166, 202, 109, 
237, 110, 82, 245, 208, 33, 232, 169, 224, 108, 32, 57, 7, 3, 139, 75, 146, 11, 130, 26, 34, 118, 40, 65, 42, 206, 225, 92, 63, 219, 238, 31, 150, 207, 181, 46, 200, 226, 85, 143, 68]

DSAC_4_64 = [106, 135, 142, 109, 226, 193, 160, 200, 81, 185, 55, 203, 71, 49, 158, 17, 164, 174, 44, 120, 83, 229, 113, 41, 134, 133, 235, 68, 130, 45, 
156, 122, 40, 144, 30, 138, 255, 191, 234, 1, 85, 137, 119, 9, 46, 95, 199, 251, 183, 75, 26, 63, 10, 177, 212, 89, 231, 8, 213, 167, 201, 225, 150, 70, 173, 243, 170, 56, 121, 42, 206, 117, 73, 146, 53, 86, 227, 195, 58, 179, 7, 126, 252, 197, 96, 181, 13, 219, 154, 52, 43, 15, 107, 20, 114, 216, 230, 147, 61, 202, 50, 162, 65, 129, 23, 33, 215, 223, 76, 116, 205, 184, 176, 87, 37, 157, 136, 159, 79, 165, 5, 172, 132, 
28, 218, 47, 57, 151, 244, 248, 180, 228, 190, 11, 88, 149, 3, 104, 124, 34, 54, 59, 224, 98, 163, 247, 186, 175, 161, 39, 4, 14, 108, 111, 24, 84, 139, 194, 127, 12, 60, 171, 64, 105, 217, 69, 48, 233, 21, 2, 99, 207, 27, 101, 91, 245, 242, 236, 128, 221, 123, 32, 82, 241, 222, 90, 72, 77, 125, 112, 31, 94, 182, 19, 253, 250, 168, 93, 36, 18, 237, 6, 66, 25, 78, 131, 51, 148, 102, 103, 67, 16, 239, 204, 220, 240, 29, 211, 35, 22, 38, 118, 153, 254, 169, 143, 62, 152, 110, 196, 155, 210, 198, 141, 214, 100, 74, 80, 145, 140, 0, 166, 189, 208, 246, 97, 178, 249, 209, 238, 187, 188, 192, 115, 92, 232]

DBIC_4_64 = [25, 226, 5, 90, 79, 177, 203, 184, 180, 75, 142, 163, 197, 237, 105, 80, 149, 178, 227, 50, 125, 48, 238, 37, 29, 69, 94, 64, 219, 61, 207, 
20, 218, 131, 71, 139, 116, 176, 247, 62, 191, 182, 249, 6, 51, 167, 34, 10, 19, 98, 21, 220, 47, 205, 159, 210, 76, 28, 211, 111, 200, 96, 251, 13, 15, 39, 108, 100, 179, 224, 82, 228, 189, 126, 115, 73, 222, 86, 166, 146, 27, 101, 112, 244, 128, 250, 8, 132, 150, 138, 54, 49, 56, 70, 104, 66, 24, 84, 145, 175, 119, 74, 234, 154, 151, 162, 87, 239, 133, 43, 212, 107, 93, 97, 57, 81, 122, 33, 161, 199, 7, 183, 241, 31, 
130, 123, 9, 158, 113, 103, 233, 254, 236, 95, 255, 88, 55, 12, 174, 240, 137, 144, 52, 23, 68, 91, 198, 187, 44, 172, 209, 26, 165, 63, 246, 168, 41, 160, 60, 3, 38, 134, 190, 152, 242, 11, 195, 223, 186, 229, 155, 188, 235, 89, 173, 110, 143, 206, 40, 141, 77, 78, 192, 169, 193, 
213, 170, 32, 164, 1, 14, 17, 156, 120, 83, 181, 217, 140, 4, 67, 127, 252, 45, 109, 58, 22, 2, 42, 136, 135, 208, 248, 99, 253, 185, 216, 196, 215, 147, 114, 148, 232, 92, 202, 0, 102, 72, 221, 46, 121, 35, 225, 18, 243, 231, 16, 106, 153, 53, 59, 230, 36, 171, 124, 245, 194, 129, 157, 204, 30, 201, 65, 85, 214, 117, 118]

NL_4_64 = [243, 15, 156, 183, 123, 189, 28, 48, 19, 197, 154, 13, 175, 26, 64, 172, 38, 195, 1, 67, 7, 62, 221, 219, 110, 181, 142, 127, 135, 32, 137, 
131, 17, 76, 2, 205, 201, 45, 80, 47, 178, 10, 254, 168, 90, 204, 162, 100, 103, 242, 24, 49, 230, 151, 79, 122, 222, 143, 121, 56, 33, 139, 
42, 83, 253, 232, 187, 212, 54, 224, 223, 155, 192, 215, 44, 89, 106, 41, 138, 216, 159, 109, 57, 220, 236, 177, 163, 98, 188, 179, 129, 95, 
61, 134, 148, 85, 53, 22, 78, 239, 126, 196, 209, 20, 72, 99, 60, 206, 31, 244, 185, 84, 104, 92, 55, 180, 174, 114, 4, 248, 169, 9, 117, 238, 237, 210, 39, 173, 75, 74, 14, 145, 153, 167, 229, 161, 225, 152, 87, 113, 119, 190, 11, 160, 81, 108, 208, 133, 116, 97, 164, 246, 202, 241, 77, 124, 191, 226, 111, 16, 82, 6, 245, 231, 35, 66, 102, 176, 194, 65, 52, 217, 18, 249, 94, 101, 147, 91, 107, 128, 171, 70, 140, 235, 43, 50, 166, 193, 146, 157, 228, 34, 112, 165, 184, 29, 12, 58, 132, 93, 105, 0, 30, 211, 233, 150, 40, 170, 88, 186, 234, 23, 8, 63, 46, 136, 68, 115, 125, 203, 213, 149, 247, 25, 227, 73, 5, 255, 252, 199, 207, 251, 120, 37, 218, 240, 158, 71, 118, 200, 3, 198, 250, 36, 144, 214, 
96, 59, 182, 27, 51, 86, 141, 130, 21, 69]

DU_Random = [145, 110, 54, 53, 133, 250, 138, 97, 216, 23, 131, 117, 248, 137, 209, 12, 101, 32, 22, 146, 239, 94, 162, 215, 62, 77, 192, 87, 136, 242, 221, 121, 8, 218, 189, 152, 90, 159, 17, 10, 217, 2, 20, 0, 238, 28, 153, 122, 124, 191, 48, 67, 83, 73, 55, 151, 228, 38, 246, 7, 50, 19, 43, 72, 168, 179, 210, 63, 26, 59, 14, 118, 190, 45, 35, 123, 44, 205, 166, 220, 167, 232, 187, 91, 229, 150, 108, 201, 98, 82, 120, 178, 171, 149, 86, 129, 140, 111, 144, 233, 173, 89, 134, 253, 193, 81, 170, 160, 80, 127, 243, 116, 181, 219, 255, 165, 70, 68, 161, 188, 241, 64, 76, 
214, 172, 16, 39, 164, 42, 61, 93, 254, 157, 196, 37, 199, 96, 143, 4, 47, 139, 175, 99, 195, 6, 222, 147, 60, 177, 71, 21, 230, 25, 128, 225, 3, 104, 112, 141, 237, 34, 231, 92, 158, 183, 115, 36, 130, 79, 1, 226, 41, 213, 180, 204, 206, 211, 135, 84, 182, 126, 107, 5, 30, 252, 66, 249, 49, 56, 236, 88, 103, 75, 202, 100, 207, 251, 223, 203, 234, 18, 208, 85, 186, 105, 24, 198, 113, 155, 156, 194, 224, 240, 148, 106, 154, 176, 142, 174, 125, 13, 57, 11, 51, 95, 74, 52, 31, 244, 114, 197, 163, 109, 9, 184, 227, 78, 46, 247, 200, 119, 58, 169, 245, 33, 29, 40, 27, 185, 65, 212, 15, 235, 132, 102, 69]

MAX_BIAS_Random = [82, 57, 15, 255, 240, 80, 184, 178, 252, 203, 73, 110, 35, 23, 168, 149, 119, 190, 7, 155, 247, 68, 200, 13, 123, 26, 140, 145, 74, 204, 118, 213, 208, 189, 79, 223, 126, 45, 235, 174, 55, 125, 67, 92, 62, 111, 128, 115, 158, 100, 76, 201, 133, 179, 32, 146, 69, 51, 180, 195, 8, 137, 199, 245, 72, 28, 253, 95, 33, 12, 3, 90, 162, 122, 153, 114, 215, 18, 2, 147, 220, 154, 20, 36, 22, 227, 27, 102, 129, 221, 134, 58, 44, 121, 159, 182, 41, 236, 186, 83, 214, 113, 248, 192, 225, 48, 52, 152, 161, 238, 241, 103, 212, 239, 11, 19, 237, 226, 202, 34, 185, 242, 144, 197, 64, 66, 176, 117, 14, 108, 40, 222, 150, 205, 132, 50, 167, 171, 175, 42, 46, 88, 4, 87, 47, 127, 183, 120, 78, 29, 138, 191, 209, 31, 206, 105, 250, 30, 139, 10, 169, 157, 112, 224, 217, 210, 218, 196, 9, 104, 94, 251, 211, 106, 136, 24, 160, 198, 148, 60, 246, 231, 164, 109, 39, 81, 97, 89, 173, 135, 244, 143, 219, 38, 249, 170, 53, 254, 172, 187, 93, 229, 194, 131, 6, 151, 61, 25, 141, 165, 163, 166, 37, 49, 
96, 71, 177, 124, 99, 5, 1, 65, 70, 75, 77, 54, 101, 16, 59, 130, 116, 243, 230, 107, 193, 98, 188, 0, 56, 233, 84, 181, 207, 86, 21, 142, 234, 63, 85, 216, 91, 232, 43, 17, 156, 228]

DSAC_Random = [6, 193, 206, 55, 109, 213, 113, 227, 203, 117, 77, 127, 200, 202, 190, 126, 171, 17, 235, 129, 239, 97, 19, 199, 252, 25, 158, 194, 179, 35, 241, 40, 33, 3, 167, 145, 112, 53, 152, 47, 34, 108, 149, 218, 229, 225, 162, 106, 189, 230, 61, 166, 253, 222, 39, 116, 240, 79, 138, 41, 249, 139, 191, 37, 96, 150, 99, 246, 195, 210, 141, 14, 119, 8, 154, 18, 133, 132, 221, 118, 237, 95, 46, 233, 170, 156, 15, 254, 180, 125, 5, 208, 122, 124, 89, 73, 103, 90, 70, 43, 243, 36, 178, 44, 29, 250, 78, 72, 214, 26, 209, 169, 205, 0, 231, 83, 157, 183, 238, 68, 215, 207, 
245, 60, 63, 20, 2, 123, 176, 62, 204, 130, 67, 86, 115, 110, 155, 76, 131, 45, 148, 23, 248, 173, 31, 69, 24, 185, 59, 13, 100, 242, 226, 32, 255, 48, 223, 146, 11, 56, 91, 159, 174, 144, 211, 153, 220, 228, 71, 75, 66, 111, 16, 80, 136, 182, 234, 192, 98, 94, 7, 87, 217, 186, 10, 64, 81, 107, 134, 82, 175, 161, 244, 143, 50, 58, 197, 147, 38, 22, 27, 198, 137, 140, 4, 164, 92, 52, 201, 251, 74, 120, 65, 42, 172, 224, 
168, 160, 216, 165, 1, 88, 188, 212, 84, 93, 177, 142, 101, 57, 236, 196, 9, 121, 151, 85, 181, 51, 54, 247, 12, 184, 104, 49, 187, 28, 105, 
163, 21, 114, 102, 135, 30, 232, 219, 128]

DBIC_Random = [75, 23, 104, 96, 106, 161, 156, 142, 48, 158, 218, 69, 150, 215, 61, 57, 8, 105, 35, 62, 100, 6, 27, 175, 170, 234, 19, 199, 85, 171, 206, 3, 184, 64, 185, 188, 89, 201, 173, 21, 67, 181, 147, 109, 86, 249, 115, 164, 237, 196, 168, 93, 153, 81, 51, 108, 174, 11, 182, 183, 72, 15, 
78, 177, 74, 122, 10, 50, 214, 131, 136, 243, 13, 208, 155, 232, 16, 140, 82, 34, 83, 187, 66, 217, 1, 33, 248, 38, 97, 47, 211, 0, 45, 252, 
152, 87, 146, 121, 193, 235, 116, 145, 254, 178, 65, 130, 210, 37, 40, 202, 118, 5, 251, 43, 79, 73, 222, 17, 22, 238, 112, 71, 7, 180, 159, 
119, 49, 63, 227, 219, 77, 226, 207, 44, 139, 221, 197, 154, 169, 255, 39, 144, 195, 60, 137, 166, 111, 117, 107, 245, 148, 229, 141, 102, 56, 113, 231, 94, 25, 125, 18, 172, 99, 241, 204, 205, 12, 223, 244, 138, 46, 167, 76, 209, 30, 90, 103, 36, 4, 32, 190, 26, 42, 133, 123, 216, 14, 151, 179, 88, 128, 236, 80, 20, 242, 68, 224, 176, 198, 55, 41, 52, 59, 2, 28, 220, 31, 200, 91, 230, 70, 127, 95, 110, 247, 92, 129, 9, 186, 84, 157, 126, 24, 165, 160, 246, 149, 134, 29, 101, 189, 233, 212, 240, 132, 143, 162, 120, 58, 191, 192, 114, 98, 250, 163, 135, 239, 
225, 253, 53, 213, 203, 54, 124, 228, 194]

NL_Random = [27, 119, 33, 126, 186, 234, 250, 46, 215, 254, 45, 176, 202, 131, 222, 245, 98, 148, 6, 163, 75, 87, 146, 191, 11, 199, 25, 83, 177, 112, 224, 3, 172, 227, 12, 71, 152, 94, 207, 201, 140, 32, 197, 101, 155, 196, 228, 123, 66, 104, 160, 23, 37, 190, 165, 170, 220, 13, 63, 108, 114, 93, 118, 19, 72, 205, 235, 240, 195, 15, 91, 243, 121, 189, 55, 232, 64, 1, 7, 88, 8, 84, 137, 200, 16, 239, 80, 68, 156, 81, 164, 43, 22, 174, 67, 229, 78, 192, 144, 225, 231, 209, 214, 26, 252, 115, 20, 4, 42, 145, 85, 53, 41, 153, 139, 143, 21, 62, 61, 216, 116, 185, 130, 39, 36, 47, 198, 251, 142, 29, 100, 217, 117, 134, 168, 74, 211, 86, 246, 54, 0, 167, 111, 218, 35, 150, 90, 161, 237, 14, 223, 133, 187, 212, 31, 183, 233, 50, 179, 28, 181, 213, 124, 166, 97, 247, 136, 9, 253, 38, 236, 159, 113, 241, 255, 109, 244, 106, 76, 44, 56, 180, 10, 203, 182, 
175, 173, 242, 103, 193, 122, 157, 238, 70, 210, 99, 5, 230, 69, 178, 79, 82, 171, 249, 120, 18, 206, 158, 132, 57, 58, 169, 52, 219, 162, 128, 96, 49, 138, 51, 125, 204, 40, 24, 208, 73, 102, 59, 141, 17, 147, 154, 77, 248, 105, 30, 95, 2, 151, 226, 188, 107, 34, 135, 65, 110, 221, 127, 60, 129, 149, 48, 92, 194, 184, 89]























\end{lstlisting}


\begin{thebibliography}{9}
\bibitem{shannon}
Claude Shannon, \emph{A Mathematical Theory of Cryptography}, 1945

\bibitem{ffields}
Gary Mullen, Daniel Panario and Claude Carlet, \emph{Handbook of Finite Fields}

\bibitem{aes}
Daemen J., Rijmen V. \emph{AES Proposal: Rijndael}, 1999

\bibitem{lucifer}
Horst Feistel. \emph{Block Cipher Cryptographic System}, US Patent 3,798,359. Filed June 30, 1971. (IBM)

\bibitem{des}
National Institute of Standards and Technology. \emph{Data Encryption Standard (DES)}, FIPS PUB 46-3, 1999

\bibitem{shamirdes}
Biham, E., Shamir, \emph{A. Differential cryptanalysis of DES-like cryptosystems}. J. Cryptology 4, 3–72 1991

\bibitem{matsui}
Matsui Mitsuru, \emph{Linear Cryptanalysis Method for DES Cipher}, 1993

\bibitem{camellia}
Kazumaro Aoki et al.,\emph{Specification of Camellia - a 125-bit Block Cipher}, Nippon Telegraph and Telephone Corporation, Mitsubishi Electric Corporation, 2001

\bibitem{serpent}
Anderson, R., Biham, E., Knudsen, L.R., \emph{Serpent: A Proposal for the Advanced Encryption Standard. NIST AES proposal}, 1998

\bibitem{present}
Bogdanov, A., Knudsen, L.R., Leander, G., Paar, C., Poschmann, A., Robshaw, M.J.B., Seurin, Y., Vikkelsoe, C., \emph{PRESENT: An Ultra-Lightweight Block Cipher} In: Paillier, P., Verbauwhede, I. (eds.) CHES 2007. LNCS, vol.4727, pp. 450–466. Springer, Heidelberg 2007

\bibitem{noekeon}
Daemen, J., Peeters, M., Van Assche, G., Rijmen, V., \emph{Nessie Proposal: the Block Cipher Noekeon}, Nessie submission, 2000, http://gro.noekeon.org/

\bibitem{ccz}
Carlet, C., Charpin, P., Zinoviev, V. . \emph{Codes, bent functions and permutations suitable for DES-like cryptosystems}, Designs, Codes and Cryptography, 15(2), 125–156, 1998

\bibitem{kangquan}
Kangquan Li and Nikolay Kaleyski, \emph{Two new infinite families of APN functions in trivariate form}, Cryptology {ePrint} Archive, Paper 2022/1522, 2022

\bibitem{nyberg}
Kaisa Nyberg, \emph{Perfect Nonlinear S-boxes}, Eurocrypt, 1991

\bibitem{nyberg2}
Kaisa Nyberg, \emph{Differentially uniform mappings for cryptography} – EUROCRYPT 1993, LNCS 765.

\bibitem{gold}
Robert Gold, \emph{Maximal recursive sequences with 3-valued recursive cross-correlation functions}. IEEE Trans. Inf. Theory, 14(1):154–156, 1968

\bibitem{kasami}
Tadao Kasami, \emph{The weight enumerators for several classes of subcodes of the 2nd order binary ReedMuller codes}, Information and Control, 18(4):369–394, 1971

\bibitem{dobbertinniho}
Hans Dobbertin, \emph{Almost perfect nonlinear power functions on GF(2n): The Niho case}, Inf. Comput., 151(1-2):57–72, 1999.

\bibitem{dobbertinwelch}
Hans Dobbertin, \emph{Almost perfect nonlinear power functions on GF($2^n$): The Welch case}, IEEE Trans. Inf. Theory, 45(4):1271–1275, 1999

\bibitem{rothaus}
O. S. Rothaus, \emph{On "Bent" functions}, Journal of Combinatorial Theory, Series A, Volume 20, Issue 3, Pages 300-305, 1976

\bibitem{hou}
X.D. Hou, \emph{Affinity of permutations of Fn 2}, Discrete Applied Mathematics 154(2) 313, 2006

\bibitem{dillon1}
K.A. Browning, J.F. Dillon, M.T. McQuistan, A.J. Wolfe. \emph{An APN permutation in dimension six} In Finite Fields: Theory and Applications- FQ9, volume 518 of Contemporary Mathematics, pages 33–42. AMS, 2010.

\bibitem{dillongen}
A. Canteaut, S. Duval and L. Perrin, \emph{A generalisation of Dillon's {APN} permutation with the best known differential and nonlinear properties for all fields of size $2^{4k+2}$}, Cryptology {ePrint} Archive, Paper 2016/887, 2016

\bibitem{carletnon}
C. Carlet, C. Ding, \emph{Highly Nonlinear Mappings} Journal of Complexity, 20(2–3): 205–244, 2004

\bibitem{canteaut}
A. Canteaut, P. Charpin, G. M. Kyureghyan, \emph{A new class of monomial bent functions}, Finite Fields and Their Applications 14 (2008) 221–241, 2007

\bibitem{feistel1}
H. Feistel, \emph{Cryptography and Computer Privacy}, Scientific American Vol. 228, No. 5, 15, 1973

\bibitem{courtois}
Nicolas Courtois, Josef Pieprzyk, \emph{Cryptanalysis of blockciphers with overdefined systems of equations} ,Proceedings of Asiacrypt 2002, LNCS 2501, pp. 267-287, Springer, 2002

\bibitem{webster}
A. F. Webster, S. E. Tavares, \emph{ON THE DESIGN OF S-BOXES},  Advances in Cryptology — CRYPTO ’85 Proceedings, LNCS,volume 218, pp 523–534, 1986

\bibitem{cid}
Martin Albrecht, Carlos Cid, \emph{Algebraic Techniques in Differential Cryptanalysis}, LNSC,volume 5665, Fast Software Encryption, pp 193–208, 2009

\bibitem{knuth}
 Knuth, Donald E. \emph{Seminumerical algorithms. The Art of Computer Programming}. Vol. 2. Reading, MA: Addison–Wesley. pp. 139–140, 1969

\bibitem{gaabouri}
El Gaabouri, I., Senhadji, M., Belkasmi, M. et al. \emph{A new S-box pattern generation based on chaotic enhanced logistic map: case of 5-bit S-box}. Cybersecurity 7, 59, 2024

\bibitem{solami}
E. A. Solami, M. Ahmad, C. Volos, M. N. Doja, M. M. S. Beg, \emph{A New Hyperchaotic System-Based Design for Efficient Bijective Substitution-Boxes}, Entropy 20(7):525, 2018

\bibitem{cui2}
Cui, J., Huang, L., Zhong, H., Chang, C., Yang, W., \emph{An improved AES S-box and its performance analysis}, International Journal of Innovative Computing, Information and Control, Volume 7, Number 5(A), pp. 22912302, 2011

\bibitem{nitaj1}
A. Nitaj, W. Susilo, J. Tonien, \emph{A New Improved AES S-box With Enhanced Properties}, 25th Australasian Conference on Information Security and Privacy (ACISP 2020), Perth, France. hal-03437913, 2020

\bibitem{mordel}
Alali, A. S., Ali, R., Jamil, M. K., Ali, J. Gulraiz. \emph{Dynamic S-Box Construction Using Mordell Elliptic Curves over Galois Field and Its Applications in Image Encryption}. Mathematics, 12(4), 587. https://doi.org/10.3390/math12040587, 2024

\bibitem{frob}
R. Ali, J. Ali, Ping Ping, M. K. Jamil, \emph{A novel S-box generator using Frobenius automorphism and its applications in image encryption
}, Nonlinear Dynamics 112(21), DOI:10.1007/s11071-024-10003-4, 2024

\bibitem{hongjun}
H. Liu, A. Kadir, C. Xu, \emph{Cryptanalysis and constructing S-Box based on chaotic map and backtracking}, Applied Mathematics and Computation 376(3):125153, DOI:10.1016/j.amc.2020.125153, 2020

\bibitem{uga}
Jean-Guillaume Dumas, Jean-Baptiste Orfila, \emph{Generating S-Boxes from Semi-fields Pseudo-extensions} [Research Report] Université Grenoble Alpes (UGA), hal-01075148, 2014

\end{thebibliography}
\end{document}